\documentclass[aps,prb,amsmath,amssymb,twocolumn,superscriptaddress,longbibliography]{revtex4-1}

\usepackage{graphicx}
\usepackage{dcolumn}
\usepackage{bm}
\usepackage{hyperref}

\usepackage{xcolor}
\usepackage{soul} 

\begin{document}

\title{Doping- and temperature-dependent electronic structure and spin dynamics in the Hubbard model on a square lattice within cluster perturbation theory}

\author{V. I. Kuz'min}
\email{kuz@iph.krasn.ru}
\affiliation{Kirensky Institute of Physics, Federal Research Center KSC SB RAS, Krasnoyarsk, 660036 Russia}

\author{S. G. Ovchinnikov}
\affiliation{Kirensky Institute of Physics, Federal Research Center KSC SB RAS, Krasnoyarsk, 660036 Russia}
\affiliation{Siberian Federal University, Krasnoyarsk, 660041 Russia}

\date{\today}

\begin{abstract}
The doping and temperature dependencies of the electronic structure and the dynamic spin susceptibility of the Hubbard model on a square lattice are studied within cluster perturbation theory (CPT). The most general features of both electronic and spin spectra qualitatively agree with the experimental data on cuprates. Generalized mean-field calculations of the electronic structure using static short-range magnetic correlations from CPT are also implemented and compared with the results of CPT, allowing us to discuss the role of short-range antiferromagnetism in the formation of the pseudogap. \end{abstract}

\maketitle

\section{\label{Intro}Introduction} 

The problem of pseudogap (PG) in cuprates is still one of the most intriguing, in spite of many important findings in both experimental and theoretical studies \cite{Timusk99,Li18,Robinson19,Vedeneev21,Kordyuk15,Pelc20,Zhu23}. Besides unconventional $d_{x^2-y^2}$-symmetry and a high value of the transition temperature $T_c$, cuprates demonstrate rather unusual normal state properties. Indeed, the problem of transformation of an undoped antiferromagnetic insulator like  $\text{La}_{2-x}\text{Sr}_x\text{CuO}_4$ into a non-magnetic Fermi liquid through the intermediate PG state remains unsolved due to strong electron correlations, which hinder the use of standard approaches of quantum theory of condensed matter. PG has been found in a range of experiments. The first was nuclear magnetic resonance (NMR) \cite{Mendels88}. Later, a number of anomalies in transport, optical measurements, Raman and inelastic neutron scattering (INS), specific heat, and angular resolved photoemission spectroscopy (ARPES) have been found (see the discussion in Ref.~\onlinecite{Kordyuk15}). In spectroscopic measurements, PG is associated with a momentum-dependent suppression of spectral weight near the Fermi level \cite{Loeser96,Hashimoto14}.

 The nature of the mysterious PG state appears to be associated with short-range antiferromagnetic correlations. Already in the early HTSC days, the scattering of conduction electrons by antiferromagnetic fluctuations was proposed to explain the PG effect. \cite{Friedel89, Alloul89, Alloul12}. Barzykin and Pines noted \cite{Barzykin95} that between the temperature $T^*$ of the maximum in the temperature dependence of the uniform spin susceptibility and the lower crossover temperature $T^{**}$ the product $^{63}T_1T$, where $^{63}T_1$ is the $\text{Cu}$ nucleus spin-lattice relaxation time, decreases linearly with temperature $T$. They showed, by considering the nearly antiferromagnetic Fermi liquid model, that this ``spin pseudogap'' behavior is related to a reduction in the density of states. To distinguish different characters of PG between the temperatures $T^{**}$ and $T^*$, discussed above, and below $T^{**}$, the terms ``weak PG'' and ``strong PG'' were introduced \cite{Schmalian98}. Two different regimes of spin fluctuation behavior with temperature were demonstrated \cite{Chubukov98, Chubukov98_1}. A large body of cluster and diagrammatic studies on the Hubbard model shows the opening of PG with the development of antiferromagnetic correlations \cite{Senechal04, Rohringer18}. Dynamic antiferromagnetic fluctuations within the PG state were shown to produce enhanced entanglement \cite{Bippus25}. Recent inelastic X-ray scattering measurements result in new insights in the problem of the electronic properties and collective spin excitations (magnons) up to high doping \cite{Zhang22}. These data reveal that spin excitations in hole-doped cuprates are related to short-ranged magnons. INS data for various families of cuprates show the types of spin excitation evolution with doping and temperature that can differ qualitatively over the PG doping range \cite{Hinkov07,Lipscombe07,Reznik08,Chan16,Chan16_1,Anderson25}, possibly indicating a nontrivial interrelation between electronic and spin spectral properties within the PG state. 

Inspired by the above findings, we plan to investigate the simultaneous evolution of the single-particle properties and spin excitations within the two-dimensional Hubbard model on a square lattice. For its theoretical description, one has to use methods that take into account the effects of short-range magnetic correlations on the electron and spin dynamics. Several approaches to treat both local and non-local correlations exist, such as diagrammatic extensions \cite{Sadovskii05,Rohringer18} of dynamical mean-field theory \cite{Metzner89,Georges96} or cluster methods \cite{Maier05}. In this paper, the doping and temperature dependence of the electronic spectral function and the dynamic spin susceptibility is studied using CPT \cite{Senechal00,Senechal02}. Our considerations are also supplemented by calculations using the simplest version of the generalized mean-field approximation (GMFA), based on the equations of motion of the Green’s functions and the Mori projection technique \cite{Mori65,Zubarev60}. The electronic spectral properties calculated within the GMFA approach are compared with CPT, which allows us to trace the influence of static short-range correlations on the electronic structure and to add simple physical interpretation to the CPT results.

In this paper, first, it is shown that the results of the CPT calculations of the doping and temperature evolutions of electronic and spin spectral functions are generally consistent with the ARPES and INS data on cuprates. Within CPT, it has been shown  that the zero-temperature doping evolution of the electronic structure, as well as its temperature evolution at some doping around optimal, qualitatively proceeds through two PG stages and depends crucially on the spin-correlation radius \cite{Kuzmin20}. By analogy to Ref.~\onlinecite{Schmalian98}, we will call the low-temperature PG variant the “strong PG”, and the other one the “weak PG”. Within the first one, the Fermi arc’s spectral weight is almost independent of doping or temperature. The latter is characterized by a smooth transition from the Fermi-arc and pronounced PG drop in the energy distribution curve (EDC) in the antinodal direction to the Fermi liquid-like state. These observations generally agree with those obtained within dynamical cluster approximation (DCA) \cite{Werner09,Gull10} and the observation of the two characteristic PG temperatures in the transport properties \cite{Pelc20}. In this paper, the evolution of the electronic structure is studied within a broad temperature range for various doping values together with static and dynamic spin correlations. Next, we discuss the comparison of the CPT data with GMFA, which leads to the conclusion that static antiferromagnetic correlations and thermal fluctuations are especially crucial for the zero-frequency PG behavior, which is the Fermi arc formation. However, taking them into account alone turns out to be completely insufficient to explain the frequency dependence of PG drop.

The remainder of this paper is structured as follows. In Sec.~\ref{sec:2}, we briefly discuss the model and methods. In Sec.~\ref{sec:3} the CPT results are presented. In Sec.~\ref{sec:4}, we discuss the results and compare them to the mean-field calculations. Section~\ref{sec:5}, contains concluding remarks. Detailed information on the calculations can be found in Appendices~\ref{sec:a} and~\ref{sec:b}.

\section{\label{sec:2} Model and methods}
We study the Hubbard model on a square lattice:
\begin{equation}
H =  - \sum\limits_{i,j,\sigma } {t_{i,j} } \hat c_{i,\sigma }^{\dag}  \hat c_{j,\sigma }^{}  + \sum\limits_i {U\hat n_{i, \uparrow } \hat n_{i, \downarrow } } ,
\end{equation}						
where the nearest and next-nearest hoppings $t$ and $t'$ are taken into account, $U$ is the local Coulomb repulsion, which is fixed as $U=8$, $\hat c_{i,\sigma }^\dag  (\hat c_{i,\sigma }^{} )
$ creates (annihilates) an electron with spin $\sigma$ at site $i$, and $\hat n_{i,\sigma }$ is the electron number operator. The calculations are performed in the paramagnetic phase.

As the main method of calculating the electronic and spin spectral function, CPT is applied here to study the electronic spectral function $A\left( {{\bf{k}},\omega } \right) =  - \frac{1}{\pi }{\mathop{\rm Im}\nolimits} G_{ \uparrow  \uparrow } \left( {{\bf{k}},\omega } \right)$ and the dynamic transverse spin susceptibility spectrum $\frac{1}{\pi }{\mathop{\rm Im}\nolimits} \chi \left( {{\bf{k}},\omega } \right)$, where $G_{ \uparrow  \uparrow }$ is the spin-up electron Green function, $\bf{k}$ is a wave vector, and $\omega$ denotes frequency. All energy quantities are measured in units of the nearest hopping integral $t$. 

Within CPT, usually, the cluster Green's function is calculated using exact diagonalization. Then, the lattice is covered by periodic translations of a cluster to form a hopping matrix $T\left(\bf{k}\right)$ and the CPT Green's function is calculated. Finally, the translation invariance is restored to obtain the spectral function. In this paper, CPT with a 12-site rectangular cluster is used. The intercluster hopping matrix is averaged among all possible periodic tilings. The resulting Green's function is averaged over the $3\times4$ and $4\times3$ clusters to recover the original lattice symmetry.

In deriving the CPT equations we follow the general approach of the Hubbard X-operator perturbation theory \cite{Zaitsev75,Ovchinnikov_book} and apply the generalized Hubbard-1 approximation for the intercluster interactions. For the spin excitations, the procedure is identical to the case of the electronic spectrum if the intercluster spin interaction is considered in the t-J model approximation. This produces the equations that are used in a random phase approximation (RPA) manner to fit the undoped spectrum to a spin-wave-like one \cite{Kuzmin24}. The details of the derivation are given in Appendix~\ref{sec:a}. 

In order to have some analytical results to compare with CPT, we use a simple GMFA variant, which is the simplest strong-coupling approximation beyond Hubbard-1 that can be obtained using the equations of motion approach \cite{Zubarev60} and Mori projection technique \cite{Mori65}. The approximation we used accounts for the dependence of the electronic spectrum on static magnetic correlations. The technical details can be found in Appendix~\ref{sec:b}.

\section{\label{sec:3} CPT results}

\begin{figure}
\includegraphics[width=1.0\linewidth]{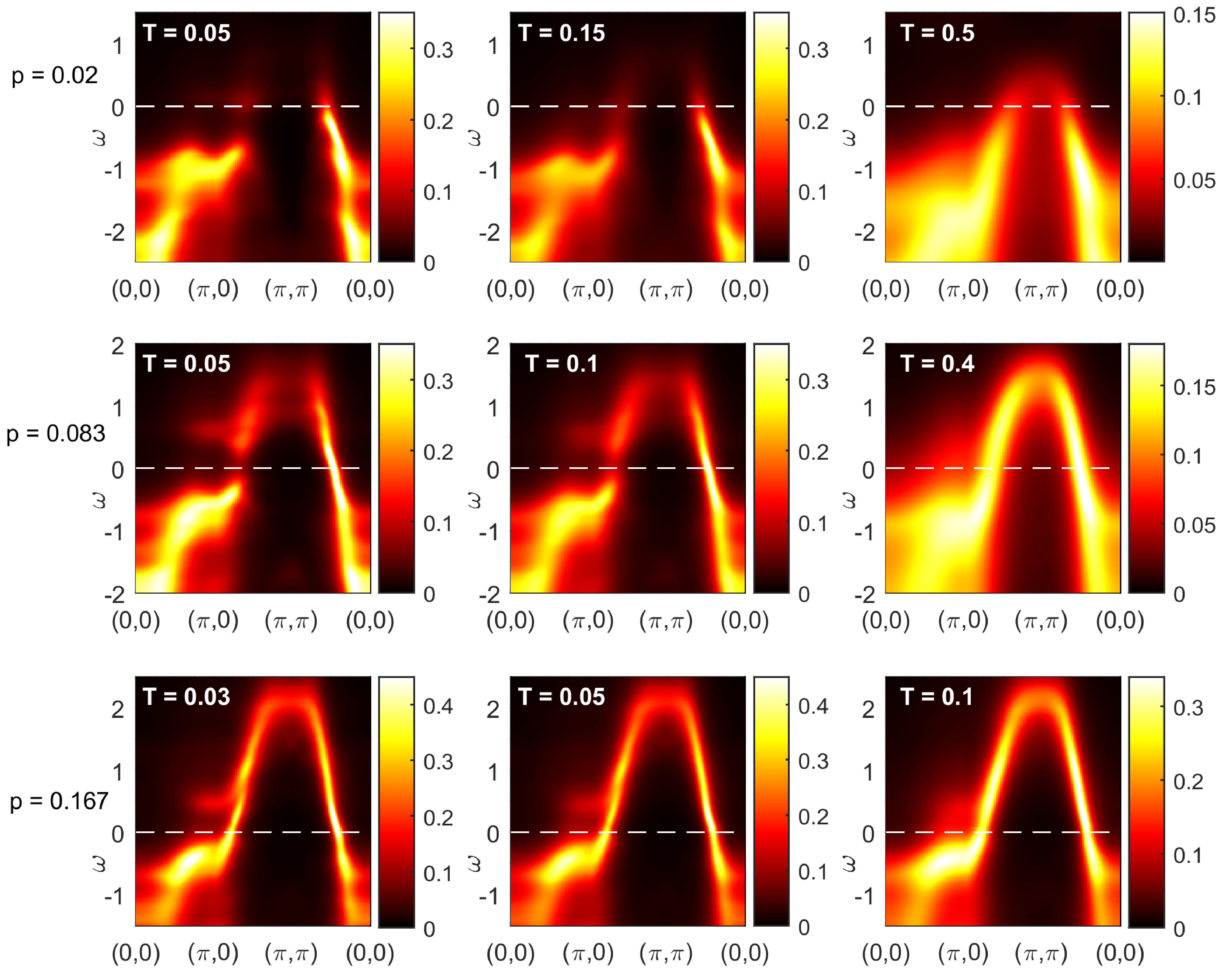}
\caption{\label{fig:A_k_w} The electronic spectral function calculated in CPT for three doping values along high-symmetry directions at various temperatures.}
\end{figure}

Figure~\ref{fig:A_k_w} provides an overview of the calculated electronic structure. For each doping level, which corresponds to a specific row in the figure, three temperatures are shown: two are relatively low, and one is high.  For two left columns of the figure (two relatively low temperatures), the spectrum is very similar around the Fermi level. At very low doping, the spectral weight in the antinodal direction close to the Fermi level at $T=0.05$ and $T=0.15$ is almost absent, while for $T=0.5$ the dispersion is pronounced in the antinodal direction almost as well as in the nodal one, and the dome-shaped signature of the free dispersion appears in the region of the $(\pi, \pi)$ point. For the underdoped case with $p=0.083$ shown in the second row of the figure, similar characteristics are observed. However, above the Fermi level, the well-defined dispersion is present, but a strong suppression of the spectral weight in the antinodal direction close to the Fermi level is observed at low temperatures. This result demonstrates the appearance of PG in the antinodal direction. The free dispersion manifests itself stronger at high temperature $T=0.4-0.5$, but, for a doping level similar to the optimal-to-high doping of cuprates in the third row, the sign of a free dispersion is already seen at the lowest temperature shown.

\begin{figure*}
\includegraphics[width=1.0\linewidth]{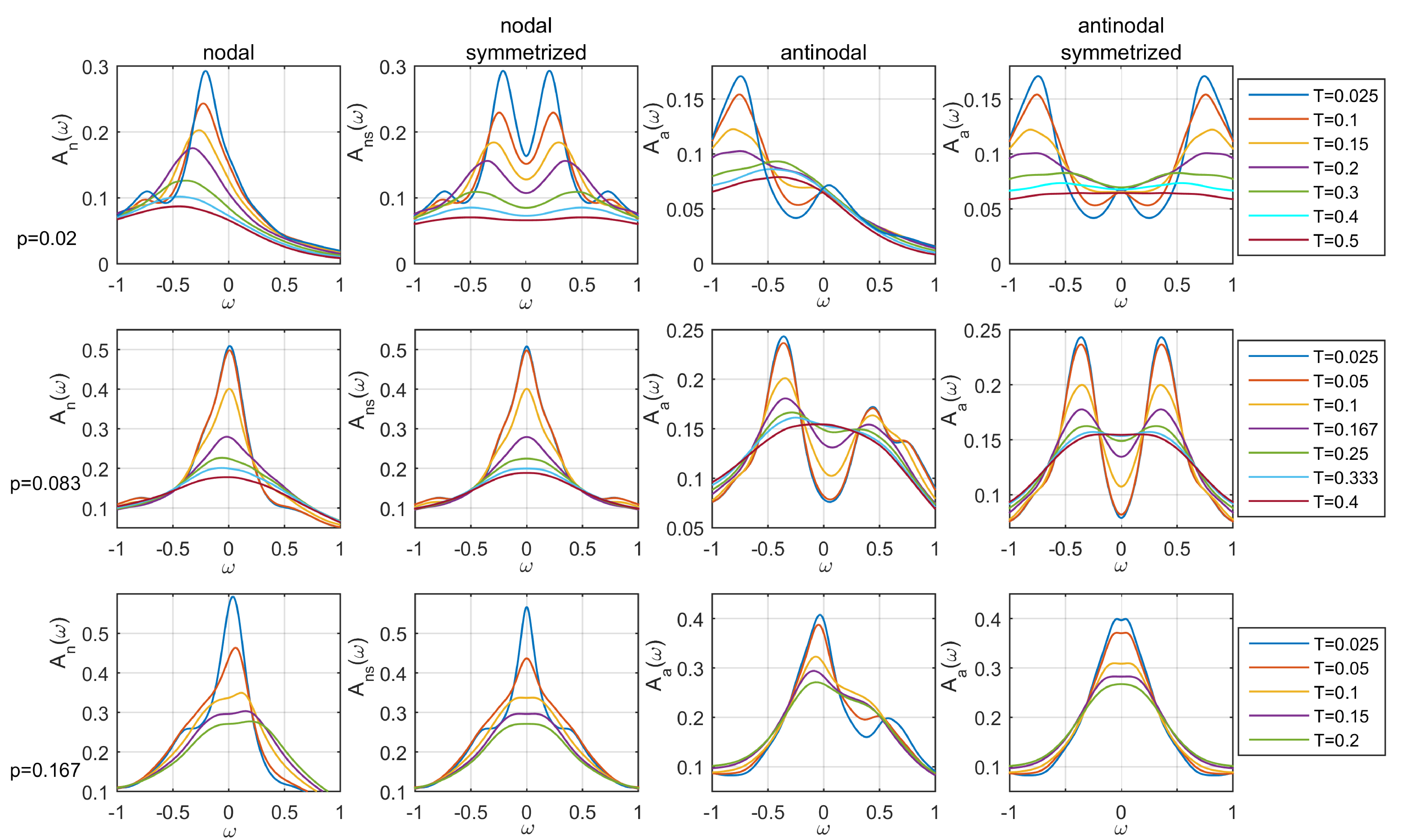}
\caption{\label{fig:A_w_all} The EDCs at the Fermi-sarface's wave vector $\mathbf{k}_F$ in the nodal and antinodal directions for the three values of doping concentration $p$ and several temperatures indicated within the insets.}
\end{figure*}

The nodal-antinodal dichotomy is a notable property of the electronic structure of the square-lattice Hubbard model and cuprates. The most noticeable signature of this dichotomy is clearly visible in regular EDCs, as seen by comparing the first and third columns of Fig~\ref{fig:A_w_all}: the antinodal spectral weight distribution has two clearly separated maxima at low temperatures, unlike the nodal one. In ARPES experiments, PG is often studied using symmetrized EDCs to visualize the presence of a gap relative to the Fermi level \cite{Hashimoto14}. In this approach, PG is defined as the drop in the symmetrized EDC at the Fermi level, and the temperature at which only one central peak remains is taken as the estimate of the location where the crossover from the normal metal to PG occurs. Technically, a gap in symmetrized EDCs can have two sources: the first is a position of the Fermi level at the edge of the peak and the second is a gap or dip in the EDC.

At very low doping $p=0.02$, the symmetrized EDC in our calculations has a gap even in the nodal direction due to the position of the chemical potential to the right of the main peak (similar positions can be found in the CPT calculations \cite{Wang20} and in near-nodal ARPES measurements \cite{Hashimoto14}). This may be a remnant of the Mott insulator state when we have a dip at the Fermi level for all directions, but it depends on momentum and is stronger for the antinodal direction. In the symmetrized EDCs, against the background of the main drop in the spectral weight, a small peak is observed at the Fermi level, which is a sharp artificial feature of the frequency symmetrization procedure at low temperatures. The dip in the symmetrized EDC in both the nodal and antinodal directions exists up to high temperatures: due to the proximity to the Mott state, the Fermi level is at the tail of the spectrum, so very high temperatures $T^* \sim  0.6$ are required to sufficiently smooth the EDCs to produce a single dome shape in the symmetrized EDC.

In the underdoped case $p=0.083$, the temperature evolution is similar to the ARPES experiments on underdoped cuprates. The symmetrized spectrum in the nodal direction has a peak decreasing in height with temperature. In the antinodal direction, there is a transition with temperature from a dip to a single dome. The EDCs in the antinodal direction have a drop at the Fermi level up to $T\sim0.3$. PG dip observed in the symmetrized EDCs closes at $T^*\sim0.4$. In the case of higher doping $p=0.167$, PG dip in the antinodal EDC is above the Fermi level and exists up to $T^*\sim0.1$. 

\begin{figure}
\includegraphics[width=1.0\linewidth]{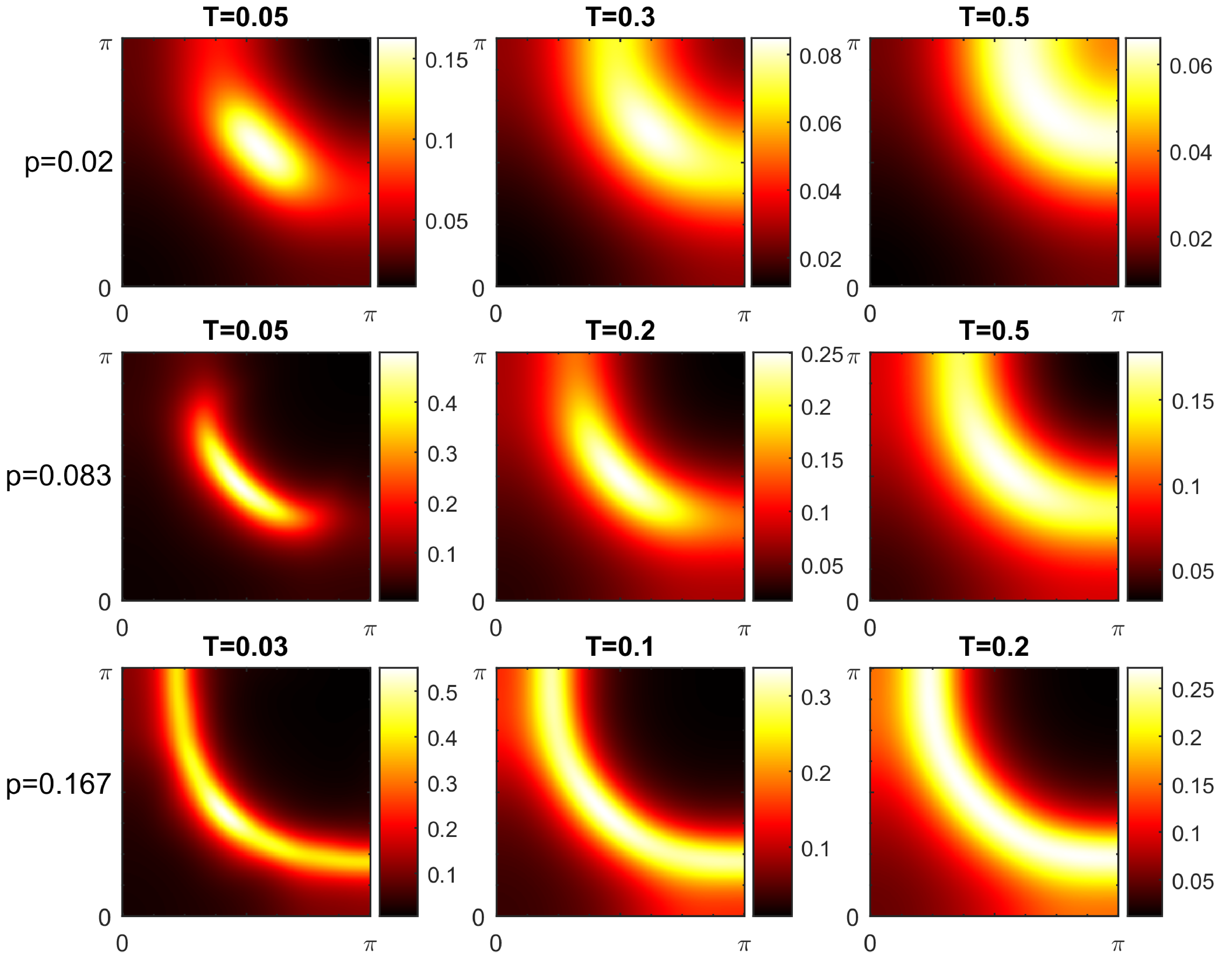}
\caption{\label{fig:Fermi} The spectral weight at the Fermi level for various doping and temperature values.}
\end{figure}

\begin{figure}
\includegraphics[width=1.0\linewidth]{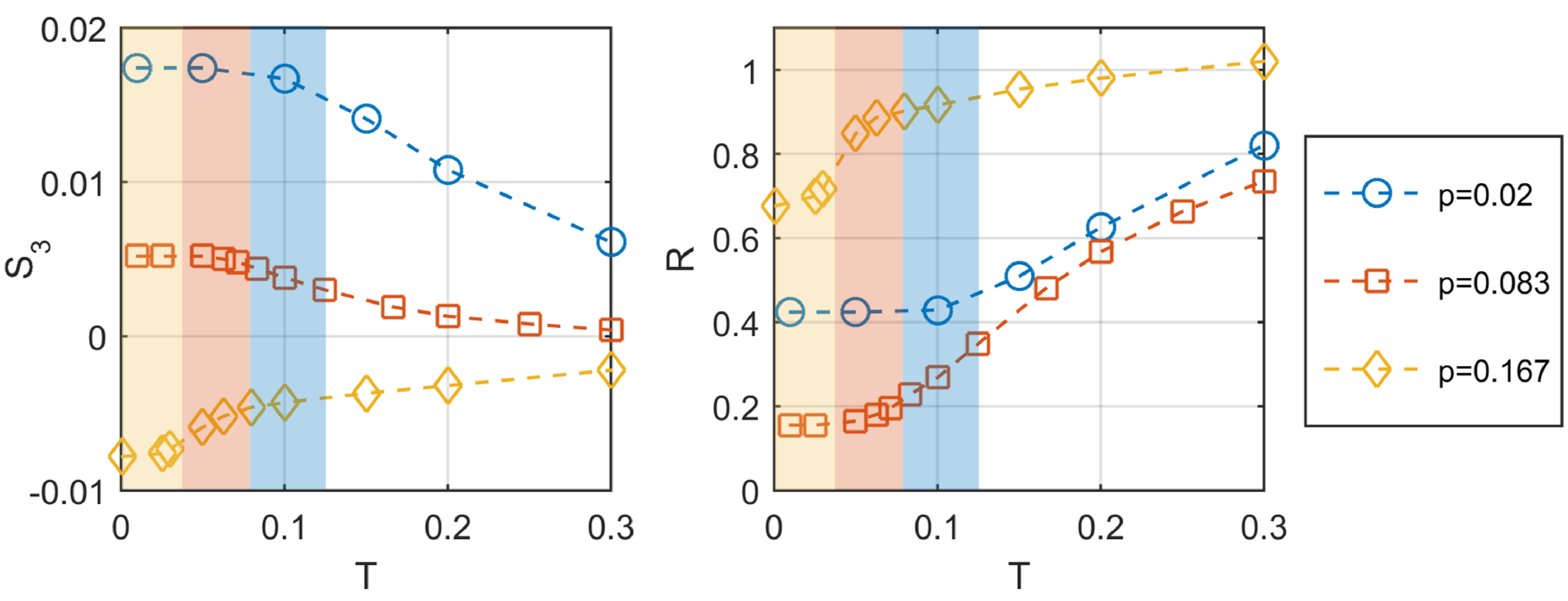}
\caption{\label{fig:RS_T} The third coordination sphere’s spin correlations (left) and the Fermi level’s antinodal to nodal ratio (right). The colored (correspondingly to the lines) regions schematically show the crossover temperature $T^{**}$, below which the Fermi arc, as well as the spin correlations, remain almost unchanged.}
\end{figure}

The spectral weight distribution at the Fermi level is shown in Fig~\ref{fig:Fermi}. Note that for each doping shown in Fig~\ref{fig:Fermi}, there exists a low-temperature scale over which the Fermi surface remains almost unchanged. This region, the low-temperature or strong PG phase, located below $T^{**}$, exists due to the stability of the ground state against thermally induced excitations. It is shown in Fig~\ref{fig:RS_T} by means of the antinodal and nodal spectral weights at the wavevector $\mathbf{k}_F$, which is qualitatively defined as a momentum-space point with maximal spectral weight in the momentum direction, and their ratio $R=A_{AN}({\bf{k}}_F)/A_{N}({\bf{k}}_F)$. Above the low-temperature PG temperature $T^{**}$, the arcs grow, forming an almost uniform spectral weight distribution of the metallic Fermi surface at high temperature, as seen in Fig~\ref{fig:Fermi}. From Fig~\ref{fig:RS_T} and Fig~\ref{fig:A_w_all}, the decrease of $T^{**}$ with doping is seen. However, no sharp boundaries between these different states are found here, only smooth crossovers. The spin correlations for the third coordinate sphere, as a qualitative measure of short-range order, are also shown in Fig~\ref{fig:RS_T}, and their behavior is similar to the temperature dependence of the spectral weight ratio.

\begin{figure}
\includegraphics[width=1.0\linewidth]{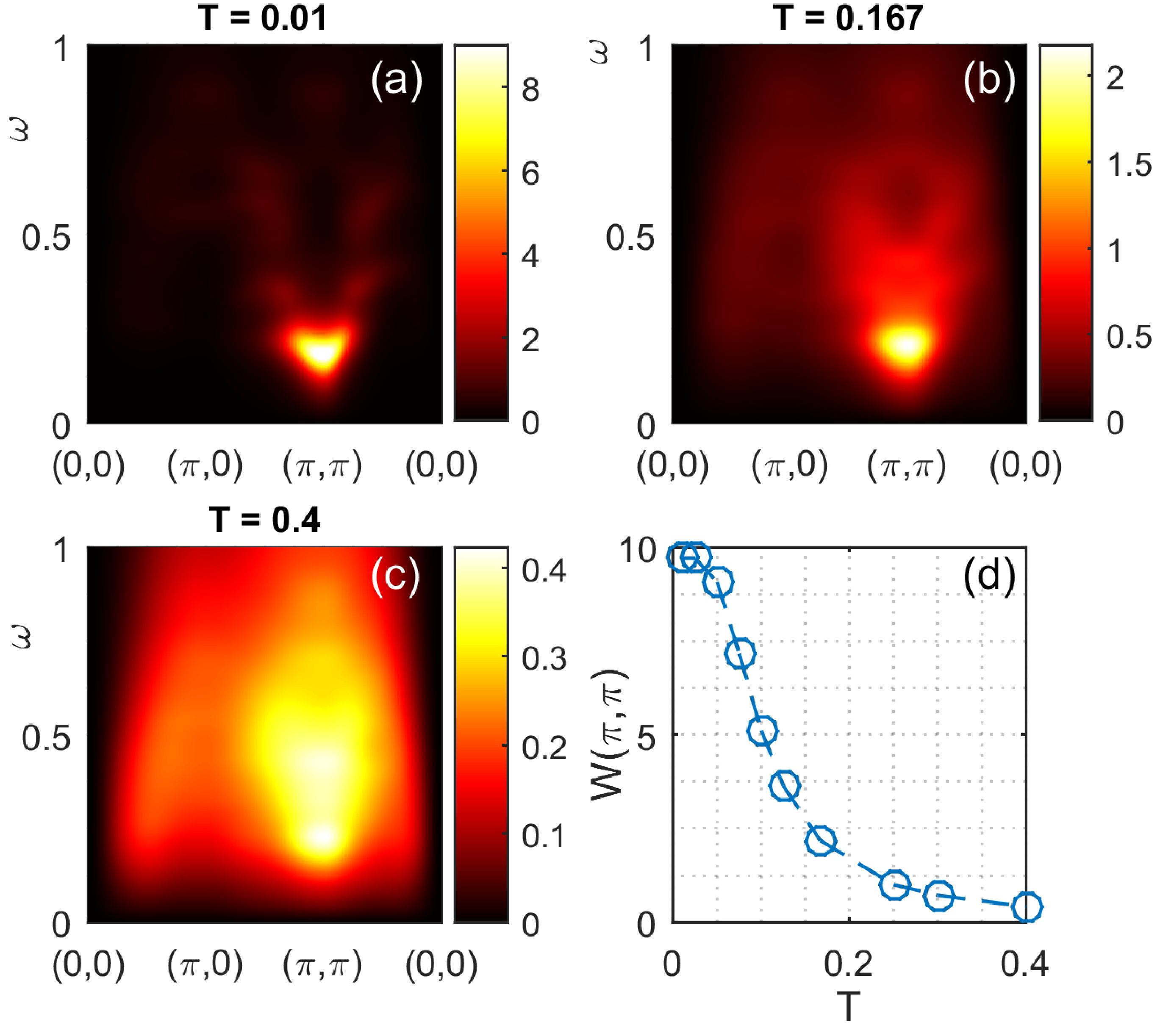}
\caption{\label{fig:chi_kw_11} (a)-(c) The CPT dynamic spin susceptibility spectra along the symmetric directions for various temperatures and (d) the temperature dependence of the maximal $\left(\pi, \pi\right)$ spectral weight for hole concentration $p=0.083$.}
\end{figure}

\begin{figure}
\includegraphics[width=1.0\linewidth]{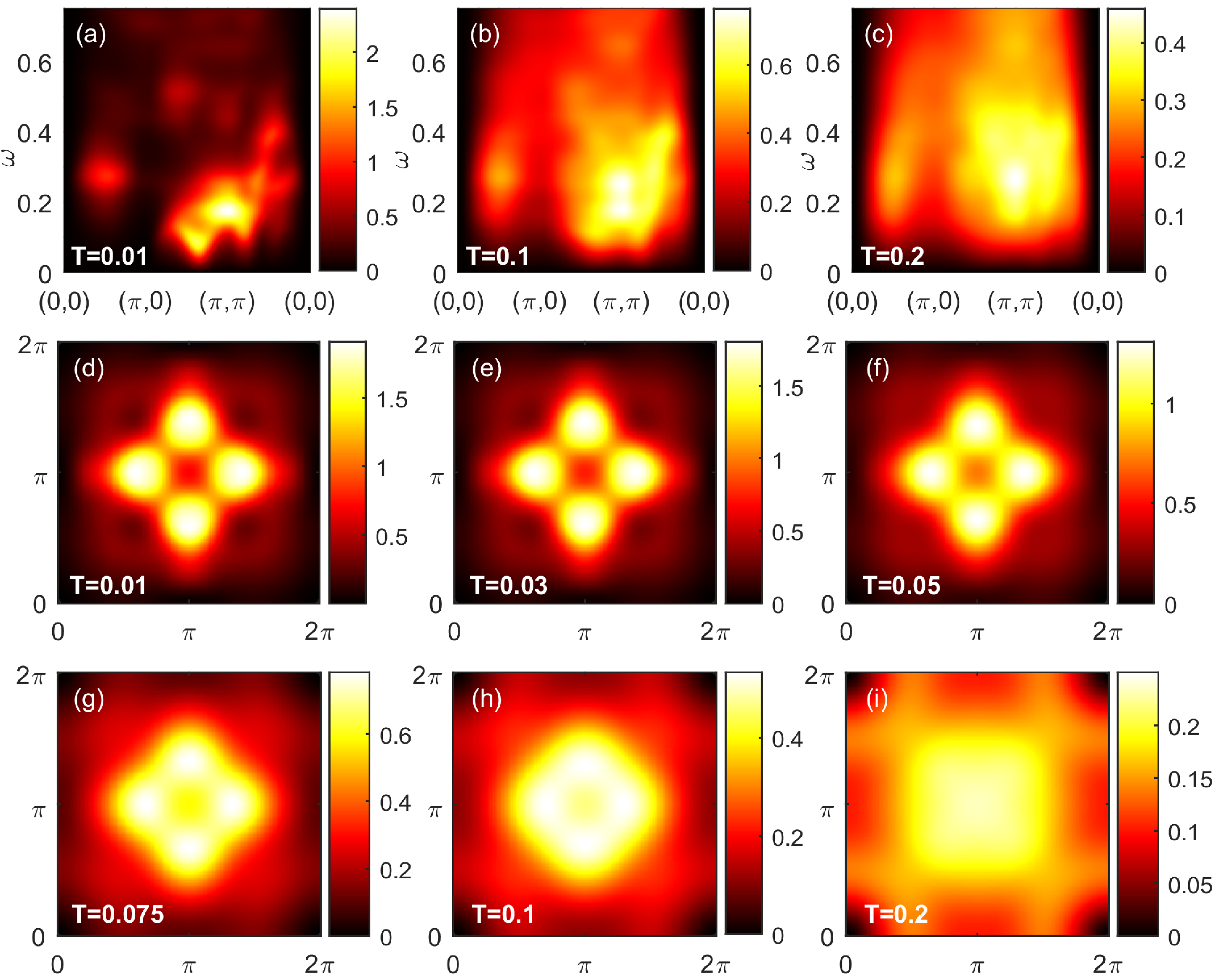}
\caption{\label{fig:chi_kw_10} The CPT dynamic spin susceptibility spectra for $p=0.167$ along symmetric directions and (d)-(i) it’s $\omega = 0.1$ constant energy cuts for various temperatures}
\end{figure}

The low-doping temperature dependence of the dynamic spin susceptibility is shown in Fig~\ref{fig:chi_kw_11} for $p=0.083$. At low temperature $T=0.01$, which is deep in the PG state, the spectrum is similar to damped spin waves, as was proposed phenomenologically \cite{Millis90, Barzykin95, Schmalian99} and obtained within GMFA and beyond \cite{Vladimirov09}. The damping of spin waves behavior qualitatively agrees with the results of Ref.~\onlinecite{Vladimirov09}, where damping near the point $(\pi, \pi)$ was shown to increase with temperature: it is seen that the spectral weight of the low-energy part smears out, which corresponds to a decrease of the correlation length. At high temperature $T=0.4$, where PG is practically absent, the spectrum resembles the weakly interacting susceptibility \cite{Kuzmin23}, but still with a significant and smeared response at $\left(\pi, \pi\right)$. The maximal $\left(\pi, \pi\right)$-point's spectral weight $W\left(\pi, \pi\right)$ exhibits a temperature dependence similar to those of the Fermi surface and short-range correlations (see Fig~\ref{fig:RS_T}) and occurs approximately in the same temperature range. At higher doping $p=0.167$, the spectral weight of the low-energy part also decreases, but the temperature scale is smaller: as shown in Fig~\ref{fig:chi_kw_10}, significant changes in the spectral weight distribution occur already at $T=0.1$. At this doping and low temperature the lowest-energy excitations in Fig~\ref{fig:chi_kw_10}(a) are incommensurate. Above them, the $(\pi, \pi)$ maximum is present. This picture is qualitatively in agreement with CPT-RPA calculations \cite{Kuzmin23}, indicating a high influence of the one-electron (renormalized) contribution to the spin susceptibility. We also suppose that such spectrum can be reproduced by an extremly frustrated damped spin wave model. Further higher in energy, the excitations in the nodal direction are mainly pronounced. This pattern is generally similar to the hourglass feature observed in many cuprates in different variations \cite{Hinkov07,Reznik08,Lipscombe07,Chan16}. The four-peak low-energy structure loses its incommensurability similarly to what happens in some cuprates \cite{Lipscombe09,Anderson25} with increasing temperature, as shown in Fig~\ref{fig:chi_kw_10}(d)-(i). The temperature range of this trend is similar to the range in which PG remnants in the electronic structure vanish, as disscussed above. Note that in some cuprates, incommensurate low-energy excitations are observed in the superconducting phase, which itself can be the cause of the incommensurate low-energy  branch \cite{Kruger07,Reznik08,Eremin12}. However, a similar response was also observed in the normal phase \cite{Zhu23}, similar to what we observed in the WPG/normal metal boundary region.  This dependence of the spin susceptibility with doping and temperature is similar to what was observed in INS experiments on cuprates \cite{Lipscombe07,Lipscombe09,Chan16,Chan16_1,Anderson25}.

\section{\label{sec:4} Comparison between CPT and GMFA results}

Having shown that the one- and two-particle properties obtained in CPT in most general terms do not contradict with ARPES and INS results, we turn to some considerations regarding the nature of PG. To further investigate the influence of short-range spin correlations, we turn to the GMFA as described in the Appendix~\ref{sec:b}. In this approximation, static spin correlations are the main ingredient of the mean-field renormalizations of the electronic dispersion.

\begin{figure}
\includegraphics[width=1.0\linewidth]{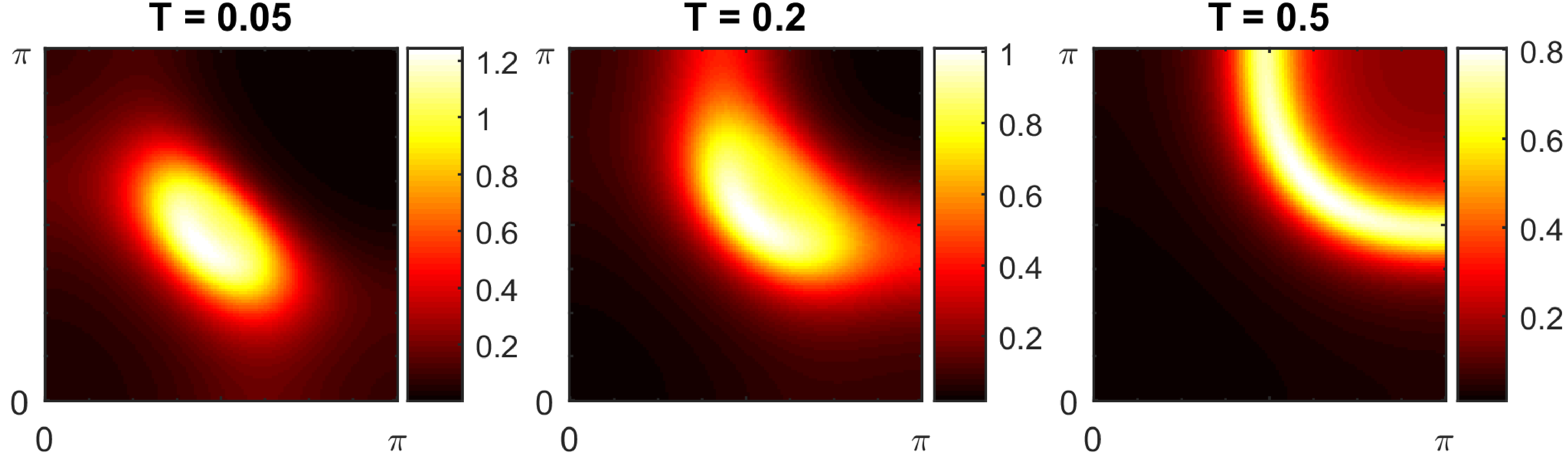}
\caption{\label{fig:GMFA_Fermi_11} (a)-(c) The temperature evolution of the GMFA Fermi surface at p=0.083.}
\end{figure}

\begin{figure}
\includegraphics[width=1.0\linewidth]{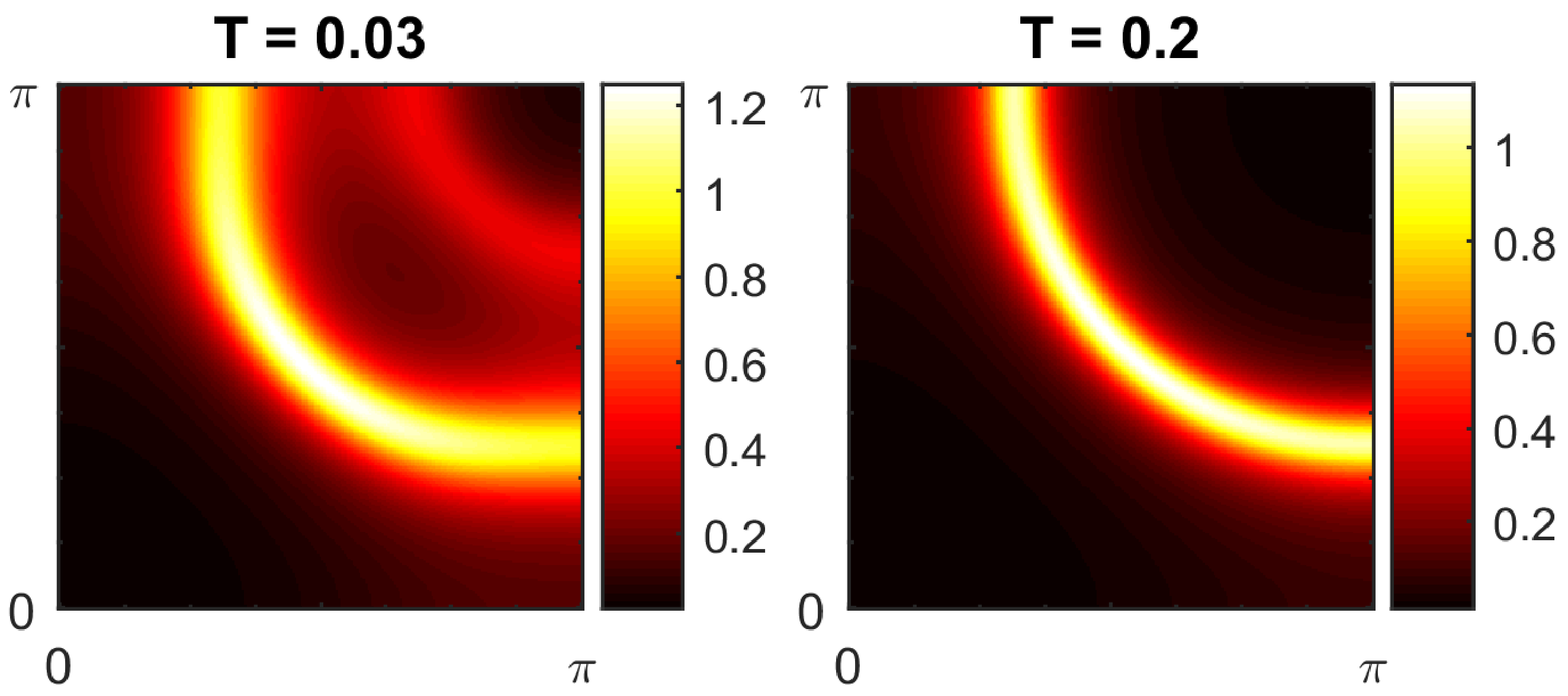}
\caption{\label{fig:GMFA_Fermi_10} The temperature evolution of the GMFA Fermi surface at p=0.167.}
\end{figure}

For the underdoped case $p = 0.083$, the evolution of the Fermi surface is shown in Fig.~\ref{fig:GMFA_Fermi_11}. It is seen that the arc-like Fermi surface gradually transforms into a Fermi surface that resembles the state of a normal metal at temperatures comparable to the corresponding temperatures in CPT (note that the same magnetic short-range order is used for the GMFA calculations as in CPT). For the higher doping case $p = 0.167$ shown in Fig.~\ref{fig:GMFA_Fermi_10}, the arc-shaped distribution is observed at low temperature $T = 0.03$, whereas at $T = 0.2$ the normal Fermi surface is already present. It should be noted that in this case the shadow Fermi surface is present near $\left(\pi, \pi\right)$, which is unavoidable for the particular implementation of GMFA, as will be discussed below. However, it is observed that taking into account the static short-range order is a crucial component of the low-energy PG physics, since it allows one to obtain the doping and temperature evolution of the Fermi surface in general accordance with CPT.

Thus, a reasonable agreement between doping and the temperature evolution of the Fermi surface is found between CPT and GMFA, provided that both approaches are supplied by a phenomenological broadening parameter (in CPT, the broadening is also necessary to obtain an interpretable electronic structure), and the latter is supplemented with band fitting parameters (see Appendix~\ref{sec:b}). However, as can be seen in Fig.~\ref{fig:compare}, in the current GMFA approximation, the electronic structure consists of only two bands, the lower and the upper, and thus cannot reproduce PG as a frequency-dependent phenomenon: note that in Fig.~\ref{fig:compare}(a), an actual gap is observed around $(\pi, \pi/2)$ in CPT, but not in GMFA. In fact, the Fermi arc distribution presented above is obtained by having a band maximum around $(\pi/2, \pi/2)$, as in Fig.~\ref{fig:compare}(a), and by using the phenomenological broadening, which transforms the Fermi pocket of the bare GMFA dispersion into an arc. As doping increases, the dispersion in Fig.~\ref{fig:compare}(c) smooths out, and the chemical potential penetrates deeper into the band, crossing the spectrum also around $(\pi, \pi/2)$, which is the cause of the shadow band presented above (note also the ``shadow'' Fermi surface in Fig.~\ref{fig:GMFA_Fermi_10}(a)). Thus, compared to the broadened CPT data as well as the recent quantum Monte Carlo results \cite{Wang25} where no evidence of back-bending dispersion was found, the pocket and the shadow Fermi surface are likely to be artificial consequences of the single-subband structure.

\begin{figure}
\includegraphics[width=1.0\linewidth]{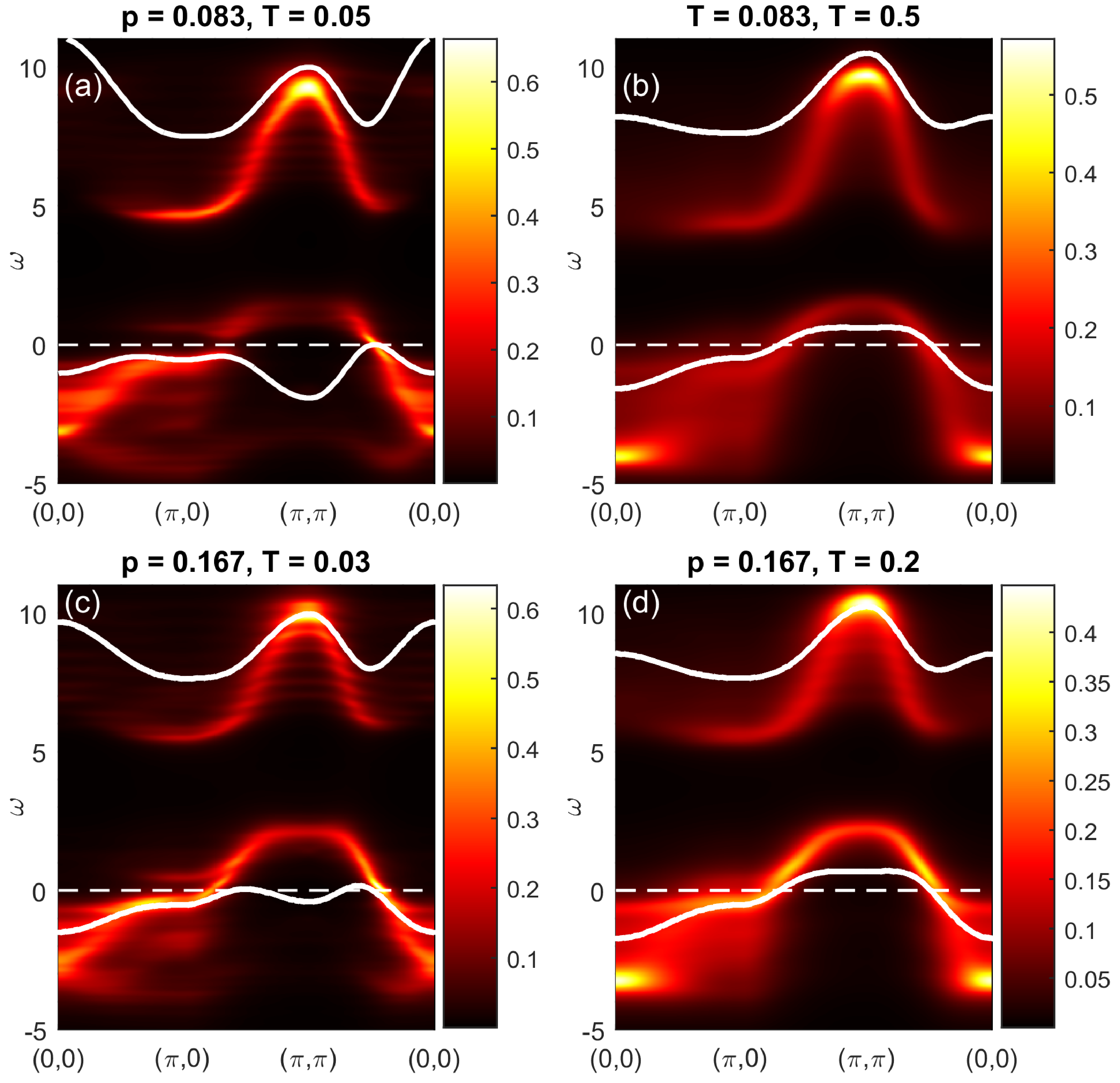}
\caption{\label{fig:compare} The CPT spectral weight shown together with the GMFA dispersion (white curves) for the two doping values and temperatures as indicated within the figure.}
\end{figure}

Finally, we discuss some issues beyond static correlations, still in terms of mean field theory and corrections beyond it. The following discussion is not supported by the electronic structure calculations explicitly tracking the influence of dynamic spin correlations in this work and is a subject for future research. As can be seen in Fig.~\ref{fig:compare}, at low doping and temperature (the strong PG state), the CPT results exhibit a significant drop in spectral intensity in the antinodal direction, similar to a gap. It has been shown that interaction with an effective fermionic excitation of a special kind can reproduce such a gap \cite{Yamaji11,Sakai16}. However, given the influence of spin correlations discussed here, the question arises whether taking into account also the dynamic spin susceptibility can have a similar effect? This can possibly be answered by taking into account the irreducible part of the self-energy beyond the GMFA. Much theoretical work has been done on this complicated problem of irreducible contributions to the self-energy \cite{Plakida07,Tung21,Belemuk16}. Of particular interest is whether extending GMFA-like calculations of the electronic structure to include the irreducible part, e.g. using the non-overlapping approximation and the dynamical susceptibility from some nonperturbative method as input, will lead to the correct PG frequency dependence.

\section{\label{sec:5} Conclusion}

In this paper, it was shown that the temperature and doping dependencies of the electronic and spin excitation spectrum are generally consistent with the ARPES and INS results. Two PG regimes, strong and weak, can exist at low and high doping and temperatures, respectively, within a wide doping range. Changes in the static and dynamic spin correlations have been shown to accompany the evolution of the electronic structure with doping and temperature.

From CPT and GMFA calculations in this paper, it is seen that the short-range static antiferromagnetic order is of a primary importance for the nodal-antinodal dichotomy and is the main ingredient of the Fermi arc formation. This can be understood in terms of dispersion renormalizations in the strong coupling limit due to static magnetic correlations. These calculations also show that thermal fluctuations, by reducing spin correlations, also reduce the dispersion renormalizations and lead to normal Fermi surfaces at some crossover temperatures, which decrease with doping. However, as is evident from comparisons of the mean-field results to CPT, dynamic correlations are likely needed in model considerations of the electronic self-energy to describe the frequency dependence of PG. The fully analytical description of PG is difficult to obtain for this reason. Moreover, for the description of cuprates, it may be important to account for the influence of phonons on spectral lines in terms of polaronic renormalizations \cite{Shneyder20}. In agreement with several previous works \cite{Gull08,Gull09}, our results imply that the PG state results from a momentum-dependent transition with two intermediate stages and occurs when the proximity of the Mott transition and strong short-range correlations are intertwined. It will be interesting to see whether the expansions of the approach applied here, when cluster calculations are used as input to the equations of motion approach, to dynamical two-particle effects and beyond, can lead to the correct description of PG frequency dependence.

\appendix
\section{\label{sec:a} Cluster perturbation theory}
The original Hamiltonian is divided into the intracluster and intercluster parts:
\begin{equation}
\hat H = \hat H_c  + \sum\limits_{f,\Delta } {\sum\limits_{a,b} {\hat c_{f,a}^\dag  T_{a,b} \left( \Delta  \right)\hat c_{f + \Delta ,b}^{} } } ,
\label{eq:1}
\end{equation}
where $f$ numbers clusters, $\Delta$ numbers their nearest neighbors, $a=\left(i,\sigma\right)$ and $b=\left(j,\sigma'\right)$ run through $L$ sites and two spin projections within a cluster, and ${\bf{T}}\left( \Delta  \right)$ is the intercluster hopping matrix.

Let us define the Hubbard operators \cite{Ovchinnikov_book} $\hat X_f^\alpha   = \hat X_f^{\left( {m,n} \right)}  = \left| m \right\rangle \left\langle n \right|$ on the eigenstates of a cluster. Annihilation operators can now be written in the following form:
\begin{equation}
\hat c_{f,a}^{}  = \sum\limits_\alpha  {\gamma _{a,\alpha } \hat X_f^\alpha  } ,
\label{eq:2}
\end{equation}
with the matrix elements being:
\begin{equation}
\gamma _{a,\alpha }  = \left\langle m \right|\hat c_a \left| n \right\rangle .
\label{eq:3}
\end{equation}

Thus, the Hamiltonian terms can be rewritten as the following:
\begin{equation}
\begin{array}{l}
 \hat H_c  = \sum\limits_p {E_p \hat X^{pp} }  \\ 
 \hat H_{cc}  = \sum\limits_{f,{\Delta }} {\sum\limits_{\alpha ,\beta } {\tilde T_{\alpha ,\beta } \left( \Delta  \right)} } \hat X_f^{\alpha ^\dag  } \hat X_{f + \Delta }^\beta  , \\ 
 \end{array}
 \label{eq:4}
\end{equation}
where $E_p$ are the cluster’s Hamiltonian eigenvalues and the hopping matrix in $X$-operator representations is:
\begin{equation}
{\tilde{\bf{ T}}}\left( \Delta  \right) = {\boldsymbol{\gamma }}^\dag  {\bf{T}}\left( \Delta  \right){\boldsymbol{\gamma }}.
 \label{eq:5}
\end{equation}

The simplest linearization can be applied to the equations of motion in terms of the generalized Hubbard-1 approximation for a partial electron annihilation excitation $\alpha$ corresponding to some nonzero matrix element of the Eqs.\ref{eq:2}-\ref{eq:3} connected to other such excitations $\beta$ by the hopping matrix ${\tilde{\bf{ T}}}\left( \Delta  \right)$:
\begin{equation}
i\frac{\partial }{{\partial t}}\hat X_f^\alpha   \approx E_\alpha  \hat X_f^\alpha   + F_\alpha  \sum\limits_\Delta  {\sum\limits_\beta  {\tilde T_{\alpha ,\beta } \left( \Delta  \right)\hat X_{f + \Delta }^\beta  } } ,
 \label{eq:6}
\end{equation}
where $E_{\alpha}=E_n-E_m$ and
\begin{equation}
F_\alpha   = n_m  + n_n 
 \label{eq:7}
\end{equation}
is the occupation factor with $n_m=<\hat{X}^{mm}>$. 

Consider the following retarded Green's function in a reduced Brillouin zone:
\begin{equation}
\begin{split}
&D_{\alpha ,\beta } \left( {{\bf{\tilde k}},\omega } \right) = \\ 
&\sum\limits_{f,g} {\int {dt} e^{i\omega t} e^{ - i{\bf{\tilde k}}\left( {{\bf{R}}_f  - {\bf{R}}_g } \right)} 
\langle\langle {\hat X_f^\alpha  (t)} | {\hat X_g^{\beta ^\dag  } (0)} \rangle\rangle } ,
\end{split}
 \label{eq:8}
\end{equation}
in order to calculate the electronic Green’s function
\begin{equation}
\begin{split}
&G_{a,b}^{CPT} \left( {{\bf{\tilde k}},\omega } \right) = 
\\ &\sum\limits_{f,g} {\int {dt} e^{i\omega t} e^{ - i{\bf{\tilde k}}\left( {{\bf{R}}_f  - {\bf{R}}_g } \right)} \langle\langle {\hat c^{}_{f,a} (t)} | {\hat c_{g,b}^\dag  (0)} \rangle\rangle } .
\end{split}
 \label{eq:9}
\end{equation}
Next, by writing the equations of motions for the two-time Green's function \cite{Zubarev60} on the right side of Eq.~\ref{eq:8}, in the approximation defined by Eq.~\ref{eq:6}, and moving to the frequency-momentum representation as in Eq.~\ref{eq:8}, one obtains:
\begin{equation}
\begin{split}
\omega D_{\alpha ,\beta } \left( {{\bf{\tilde k}},\omega } \right) = &\delta _{\alpha ,\beta }F_\alpha   + E_\alpha  D_{\alpha ,\beta } \left( {{\bf{\tilde k}},\omega } \right) + \\
&F_\alpha  \sum\limits_{\beta '} {\tilde T_{\alpha ,\beta '} \left( {{\bf{\tilde k}}} \right)} D_{\beta ',\beta } \left( {{\bf{\tilde k}},\omega } \right).
\end{split}
 \label{eq:10}
\end{equation}
From Eq.~\ref{eq:10}, it is seen that the cluster Green’s function is the following:
\begin{equation}
D_{\alpha ,\beta }^{(0)} \left( \omega  \right) = \frac{{\delta _{\alpha \beta } F_\alpha  }}{{\omega  - E_\alpha  }}.
 \label{eq:11}
\end{equation}
Thus, Eq.~\ref{eq:10} can be written as:
\begin{equation}
{\bf{D}}\left( {{\bf{\tilde k}},\omega } \right) = {\bf{D}}^{(0)} \left( \omega  \right) + {\bf{D}}^{(0)} \left( \omega  \right){\bf{\tilde T}}\left( {{\bf{\tilde k}}} \right){\bf{D}}\left( {{\bf{\tilde k}},\omega } \right).
 \label{eq:12}
\end{equation}
Since the electronic Green’s function of a superlattice is:
\begin{equation}
{\bf{G}}^{CPT} \left( {{\bf{\tilde k}},\omega } \right) = {\boldsymbol{\gamma D}}\left( {{\bf{\tilde k}},\omega } \right){\boldsymbol{\gamma }}^\dag  ,
 \label{eq:13}
\end{equation}
one obtains the CPT approximation to Eq.~\ref{eq:9}:
\begin{equation}
{\bf{G}}^{CPT} \left( {{\bf{\tilde k}},\omega } \right) = {\bf{G}}^{(0)} \left( \omega  \right) + {\bf{G}}^{(0)} \left( \omega  \right){\bf{T}}\left( {{\bf{\tilde k}}} \right){\bf{G}}^{CPT} \left( {{\bf{\tilde k}},\omega } \right),
 \label{eq:14}
\end{equation}
where $\bf{G}^{(0)}$ is the electronic Green’s function of a cluster.
Then, the lattice Green’s function can be approximated in the following way to restore the translational invariance \cite{Senechal00}:
\begin{equation}
G_{\sigma ,\sigma '} \left( {{\bf{k}},\omega } \right) = \frac{1}{L}\sum\limits_{i,j}^L {G_{(i,\sigma ),(j,\sigma ')}^{CPT} \left( {{\bf{k}},\omega } \right)e^{ - i{\bf{k}}\left( {{\bf{r}}_i  - {\bf{r}}_j } \right)} } .
 \label{eq:15}
\end{equation}

In our calculations, we work with a fixed number of particles. When there is a need for a particular intermediate doping level, which does not correspond to an integer number of particles in a cluster, the following approximation for the occupation numbers $n_m$ in Eq.~\ref{eq:7} is assumed: To obtain the desired electron density $n_e$, we set a number of electrons per cluster by assuming the occupations $1-x$ and $x$ for only two sectors of the Hilbert space, with $N$ and $N-1$ electrons:
\begin{equation}
n_e  = (1 - x)N + x(N - 1).
 \label{eq:16}
\end{equation}
In this way, the occupation numbers are supposed to be calculated as:
\begin{equation}
\begin{array}{l}
 n_{m \in \left\{ {N - 1} \right\}}  = \frac{x}{{Z_{N - 1} }}{\mathop{\rm e}\nolimits} ^{ - \beta E_m }  \\ 
 n_{m \in \left\{ N \right\}}  = \frac{{1 - x}}{{Z_N }}{\mathop{\rm e}\nolimits} ^{ - \beta E_m } . \\ 
 \end{array}
 \label{eq:17}
\end{equation}
For the electronic Green’s function, the procedure is simply equivalent to:
\begin{equation}
{\bf{G}}^{(0)} \left( \omega  \right) = x{\bf{G}}^{(0,N - 1)} \left( \omega  \right) + (1 - x){\bf{G}}^{(0,N)} \left( \omega  \right),
 \label{eq:18}
\end{equation}
where ${\bf{G}}^{(0,N - 1)} \left( \omega  \right)$ and ${\bf{G}}^{(0,N)} \left( \omega  \right)$ are the cluster Green’s functions for sectors with $N-1$ and $N$ electrons. 

In practice, our calculations are done as follows. ${\bf{G}}^{(0,N - 1)} \left( \omega  \right)$ and ${\bf{G}}^{(0,N)} \left( \omega  \right)$ are calculated using the block Lanczos method \cite{Seki18} to obtain ${\bf{G}}^{(0)} \left( \omega  \right)$. Then, CPT is applied and the lattice Green’s function and its spectral function are obtained.

In the case of the spin excitation spectrum, in order to build a similar scheme in the regime of strong electronic correlations, we roughly suppose that the intercluster bonds are described by the t-J model \cite{Chao77}:
\begin{equation}
\begin{split}    
&\hat H_{cc}^{tJ}  = \sum\limits_{f,\Delta } {\sum\limits_{a,b} {\hat a_{f,a}^\dag  T_{a,b} \left( \Delta  \right)\hat a_{f + \Delta ,b}^{} } }  + \\
&\sum\limits_{f,\Delta } {\sum\limits_{i,j} {\alpha _J J_{i,j} \left( \Delta  \right)\left( {{\bf{\hat S}}_{f,i} {\bf{\hat S}}_j  - \frac{{\hat n_{i, \uparrow } \hat n_{j \downarrow } }}{4}} \right)} } ,
\end{split}
 \label{eq:19}
\end{equation}
where $\hat a_{f,a}^\dag$ and $\hat a_{f + \Delta ,b}^{}$ are the constrained electron operators, and $J_{i,j} \left( \Delta  \right)$ is the intercluster exchange matrix:
\begin{equation}
J_{i,j} \left( \Delta  \right) = 2T_{\left( {i,\sigma } \right),\left( {j,\sigma } \right)}^2 \left( \Delta  \right)/U
 \label{eq:20}
\end{equation}
In Eq.~\ref{eq:19}, the $\alpha _J$-factor is introduced by hand to treat the influence of the interaction in an RPA manner: its value corresponding to the susceptibility instability at half filling and zero temperature is used to fit the spectrum to a spin-wavelike form.

From this point on, the entire procedure described above for the electronic spectrum is repeated to obtain the Green’s function for the partial spin excitation $\lambda$ of the spin ladder operator:
\begin{equation}
\hat S_{f,i}^ +   = \sum\limits_\lambda  {\gamma ^ \bot  _{i,\lambda } \hat X_f^\lambda  } .
 \label{eq:21}
\end{equation}
For excitations of this kind, the generalized Hubbard-1 equation is the following:
\begin{equation}
{\bf{D}}^ \bot  \left( {{\bf{\tilde k}},\omega } \right) = {\bf{D}}^{ \bot (0)} \left( \omega  \right) + {\bf{D}}^{ \bot (0)} \left( \omega  \right){\bf{\tilde J}}^ \bot  \left( {{\bf{\tilde k}}} \right){\bf{D}}\left( {{\bf{\tilde k}},\omega } \right),
 \label{eq:22}
\end{equation}
where ${\bf{\tilde J}}^ \bot  \left( \Delta  \right) = {\boldsymbol{\gamma }}^{ \bot ^\dag  } {\bf{J}}\left( \Delta  \right){\boldsymbol{\gamma }}^ \bot .$

Thus, the transverse retarded susceptibility,
\begin{equation}
\begin{split}
&{\bf{\chi }}_{i,j}^{CPT} \left( {{\bf{\tilde k}},\omega } \right) = 
\\ &\sum\limits_{i,j} {\int {dt} e^{i\omega t} e^{ - i{\bf{\tilde k}}\left( {{\bf{R}}_f  - {\bf{R}}_g } \right)} \langle\langle {S_{f,i}^ +  (t)} | {S_{g,j}^ -  (0)} \rangle\rangle } ,
\end{split}
 \label{eq:23}
\end{equation}
is obtained in the following way:
\begin{equation}
{\bf{\chi }}^{CPT} \left( {{\bf{\tilde k}},\omega } \right) = {\bf{\chi }}^{(0)} \left( \omega  \right) + {\bf{\chi }}^{(0)} \left( \omega  \right){\bf{J}}\left( {{\bf{\tilde k}}} \right){\bf{\chi }}^{CPT} \left( {{\bf{\tilde k}},\omega } \right),
 \label{eq:24}
\end{equation}
where ${\bf{\chi }}^{(0)} \left( \omega  \right)$ is the transverse susceptibility of a cluster calculated from exact diagonalization to give us a strong-coupling analog of RPA. The numerical scheme is analogous to the case of the electronic Green’s function. After restoring translation invariance, the value of $\alpha _J$ is found in the way described above and used to obtain the imaginary part of the transverse dynamical susceptibility. The electronic and spin spectral functions are calculated with Lorentzian broadenings $\delta=0.16t$ and $\delta=0.05t$, respectively.

\section{\label{sec:b} Generalized mean-field approximation}
Here we briefly discuss the implementation of the so-called generalized mean-field approximation (GMFA) in the paramagnetic phase. Given the eigenstates of the one-site Hubbard model denoted as $\{0, \uparrow, \downarrow,2\}$, one has the expression:
\begin{equation}
\hat c_{i, \uparrow }^\dag   = \hat X_{i, \uparrow }^{ \uparrow ,0}  + \hat X_{i, \uparrow }^{2, \downarrow } .
 \label{eq:1b}
\end{equation}
The normal $X$-operator Green function is defined as the matrix with an element of the form:
\begin{equation}
D_{i,j}^{GMFA}  = \left\langle {\left\langle {\left. {{\bf{\hat \Psi }}_i } \right|{\bf{\hat \Psi }}_j^\dag  } \right\rangle } \right\rangle ,
 \label{eq:2b}
\end{equation}
where
\begin{equation}
{\bf{\hat \Psi }}_j^\dag   = \left( {\hat X_j^{ \uparrow ,0} ,\hat X_j^{2, \downarrow } } \right).
 \label{eq:3b}
\end{equation}

Using the equations of motion approach \cite{Zubarev60} and Mori projection technique \cite{Mori65}, within the GMFA, the spectrum $E_{i,l}$ can be obtained from the following linearization condition (see, for example, Refs.~\onlinecite{Plakida07,Plakida13,Korshunov07,Shneyder20}, where more rigorous approximations were developed):
\begin{equation}
\left\langle {\left[ {\left[ {\hat \Psi _i ,\hat H} \right] - \sum\limits_l {E_{i,l} } \hat \Psi _l ,\hat \Psi _j^\dag  } \right]_ +  } \right\rangle  = 0.
 \label{eq:4b}
\end{equation}
where $\bf{I}$ is the identity matrix (the space index is omitted in the paramagnetic phase), and ${\bf{Q}} = \left\langle {\left. {{\bf{\hat \Psi }}_i } \right|{\bf{\hat \Psi }}_i^\dag  } \right\rangle$ is equal to:
\begin{equation}
{\bf{Q}} = \left( {\begin{array}{*{20}c}
   {Q_1 } & 0  \\
   0 & {Q_2 }  \\
\end{array}} \right) = \left( {\begin{array}{*{20}c}
   {1 + p} & 0  \\
   0 & {1 - p}  \\
\end{array}} \right)/2,
 \label{eq5b}
\end{equation}
where $p$ stands for doping concentration. The parameters $\alpha_{t'}$ and $\alpha_c$ defined below are introduced here by hand to fit the GMFA spectrum for a better comparison with CPT. The ``clear'' approximation, without them, is defined by $\alpha_{t'}=0$ and $\alpha_c=1$. In this paper, we use $\alpha_{t'}=-0.1$ and $\alpha_c=1.75$. The energy spectrum is defined as the following if we assume that $\langle \hat{n}_i \hat{n}_j \rangle = \langle \hat{n}_i \rangle \langle \hat{n}_j \rangle$ for simplicity:
\begin{equation}
\begin{array}{*{20}c}
   {\hat E(\alpha _{t'} ,\alpha _c ) = \left( {\begin{array}{*{20}c}
   {\varepsilon _{11} \left( {{\bf{k}},\alpha _{t'} ,\alpha _c \,} \right)} & {\varepsilon _{12} \left( {{\bf{k}},\alpha _{t'} ,\alpha _c \,} \right)}  \\
   {\varepsilon _{21} \left( {{\bf{k}}\,,\alpha _{t'} ,\alpha _c } \right)} & {\varepsilon _{22} \left( {{\bf{k}}\,,\alpha _{t'} ,\alpha _c } \right)}  \\
\end{array}} \right),}  \\
   {\varepsilon _{11} \left( {{\bf{k}}\,} \right) = Q_1^2 t\left( {{\bf{k}},\alpha _{t'} } \right) + \frac{{\alpha _c }}{N}\sum\limits_{\bf{q}} {t\left( {{\bf{k}}\, - \,{\bf{q}},\alpha _{t'} } \right)S\left( {\bf{q}} \right)} }  \\
   {\varepsilon _{12} \left( {{\bf{k}}\,} \right) = Q_1 Q_2 t\left( {{\bf{k}},\alpha _{t'} } \right) - \frac{{\alpha _c }}{N}\sum\limits_{\bf{q}} {t\left( {{\bf{k}}\, - \,{\bf{q}},\alpha _{t'} } \right)S\left( {\bf{q}} \right)} }  \\
   {\varepsilon _{22} \left( {{\bf{k}}\,} \right) = Q_2 U + Q_2^2 t\left( {{\bf{k}},\alpha _{t'} } \right) + \frac{{\alpha _c }}{N}\sum\limits_{\bf{q}} {t\left( {{\bf{k}}\, - \,{\bf{q}},\alpha _{t'} } \right)S\left( {\bf{q}} \right)} ,}  \\
\end{array}
 \label{eq6b}
\end{equation}
where $\varepsilon _{12} \left( {{\bf{k}}\,} \right) = \varepsilon _{21} \left( {{\bf{k}}\,} \right)$.

Thus, the equation for the Green function that we use is the following:
\begin{equation}
{\bf{D}}^{GMFA} \left( {{\bf{k}},\omega } \right) = {\bf{Q}}{\rm{[}}\omega {\bf{I}} - {\bf{E}}\left( {{\bf{k}},\alpha _{t'} ,\alpha _c } \right)]^{ - 1} ,
 \label{eq:7b}
\end{equation}
where $S(\bf{q})$ is a spin structure factor, which is estimated from the exact diagonalization for the same parameters as in CPT, so the parameter $\alpha_c$ adjusts the influence of spin correlations, while $\alpha_{t'}$ adjusts next-nearest hopping:

\begin{equation}
t\left( {\bf{k}} \right) = 2t\left[ {\cos \left( {k_x } \right) + \cos \left( {k_y } \right)} \right] +  4\left( {t' - \alpha _{t'} } \right) {\cos \left( k_x \right) \cos \left(k_y\right)} .
 \label{eq:8b}
\end{equation}
Finally, to plot the GMFA electronic structure, the same Lorentzian broadening as for CPT is used.

\begin{acknowledgments}
Sections \ref{Intro}, \ref{sec:4}, and Appendix~\ref{sec:b} have been carried out along the state assignment of the Kirensky Institute of Physics, sections \ref{sec:2}, \ref{sec:3}, \ref{sec:5}, and Appendix~\ref{sec:a} - with the support of the Russian Science Foundation, project no. 24-12-00044.
\end{acknowledgments}

\bibliography{refs.bib}

@article{Li18,
  title = {Coherent organization of electronic correlations as a mechanism to enhance and stabilize high-{TC} cuprate superconductivity},
  author = {Li, Haoxiang and Zhou, Xiaoqing and Parham, Stephen and Reber, Theodore J. and Berger, Helmuth and Arnold, Gerald B. and Dessau, Daniel S.},
  journal = {Nature Communications},
  volume = {9},
  number = {1},
  pages = {26},
  year = {2018},
  month = {Jan},
  publisher = {Nature Publishing Group},
  doi = {10.1038/s41467-017-02422-2},
  url = {https://www.nature.com}
}

@article{Vedeneev21,
  title = {Pseudogap problem in high-temperature superconductors},
  author = {Vedeneev, Sergei I.},
  journal = {Physics-Uspekhi},
  volume = {64},
  number = {9},
  pages = {890},
  year = {2021},
  publisher = {IOP Publishing},
  doi = {10.3367/UFNe.2020.12.038896},
  url = {https://iopscience.iop.org/article/10.3367/UFNe.2020.12.038896}
}

@article{Hashimoto14,
  title = {Energy gaps in high-transition-temperature cuprate superconductors},
  author = {Hashimoto, Makoto and Vishik, Inna M. and He, Rui-Hua and Devereaux, Thomas P. and Shen, Zhi-Xun},
  journal = {Nature Physics},
  volume = {10},
  number = {7},
  pages = {483},
  year = {2014},
  month = {Jul},
  publisher = {Nature Publishing Group},
  doi = {10.1038/nphys3009},
  url = {https://www.nature.com}
}

@article{Bippus25,
  title = {Entanglement in the pseudogap regime of cuprate superconductors},
  author = {Bippus, Frederic and Krsnik, Juraj and Kitatani, Motoharu and Ak\ifmmode \check{s}\else \v{s}\fi{}amovi\ifmmode \acute{c}\else \'{c}\fi{}, Luka and Kauch, Anna and Bari\ifmmode \check{s}\else \v{s}\fi{}i\ifmmode \acute{c}\else \'{c}\fi{}, Neven and Held, Karsten},
  journal = {Phys. Rev. B},
  volume = {112},
  issue = {8},
  pages = {L081110},
  numpages = {7},
  year = {2025},
  month = {Aug},
  publisher = {American Physical Society},
  doi = {10.1103/xk42-b9cx},
  url = {https://link.aps.org/doi/10.1103/xk42-b9cx}
}

@article{Zhang22,
  title = {Unraveling the nature of spin excitations disentangled from charge contributions in a doped cuprate superconductor},
  author = {Zhang, Wenliang and Agrapidis, Cli{\`o} Efthimia and Tseng, Yi and Asmara, Teguh Citra and Paris, Eugenio and Strocov, Vladimir N. and Giannini, Enrico and Nishimoto, Satoshi and Wohlfeld, Krzysztof and Pelliciari, Riccardo and Schmitt, Thorsten},
  journal = {npj Quantum Materials},
  volume = {7},
  number = {1},
  pages = {123},
  year = {2022},
  month = {Dec},
  publisher = {Nature Publishing Group},
  doi = {10.1038/s41535-022-00528-5},
  url = {https://www.nature.com/articles/s41535-022-00528-5}
}

@article{Lipscombe07,
  title = {Persistence of High-Frequency Spin Fluctuations in Overdoped Superconducting {$La_{2-x}Sr_xCuO_4$} ({$x=0.22$})},
  author = {Lipscombe, O. J. and Hayden, S. M. and Vignolle, B. and McMorrow, D. F. and Perring, T. G.},
  journal = {Physical Review Letters},
  volume = {99},
  number = {6},
  pages = {067002},
  year = {2007},
  month = {Aug},
  publisher = {American Physical Society},
  doi = {10.1103/PhysRevLett.99.067002},
  url = {https://link.aps.org/doi/10.1103/PhysRevLett.99.067002}
}

@article{Lipscombe09,
  title = {Emergence of Coherent Magnetic Excitations in the High Temperature Underdoped {${\mathrm{La}}_{2\ensuremath{-}x}{\mathrm{Sr}}_{x}{\mathrm{CuO}}_{4}$} Superconductor at Low Temperatures},
  author = {Lipscombe, O. J. and Vignolle, B. and Perring, T. G. and Frost, C. D. and Hayden, S. M.},
  journal = {Phys. Rev. Lett.},
  volume = {102},
  issue = {16},
  pages = {167002},
  numpages = {4},
  year = {2009},
  month = {Apr},
  publisher = {American Physical Society},
  doi = {10.1103/PhysRevLett.102.167002},
  url = {https://link.aps.org/doi/10.1103/PhysRevLett.102.167002}
}

@article{Anderson25,
  title = {Gapped commensurate antiferromagnetic response in a strongly underdoped model cuprate superconductor},
  author = {Anderson, Zachary W. and Tang, Yang and Nagarajan, Vikram and Chan, Mun K. and Dorow, Chelsey J. and Yu, Guichuan and Abernathy, Douglas L. and Christianson, Andrew D. and Mangin-Thro, Lucile and Steffens, Paul and Sterling, Tyler and Reznik, Dmitry and Bounoua, Dalila and Sidis, Yvan and Bourges, Philippe and Greven, Martin},
  journal = {npj Quantum Materials},
  volume = {10},
  number = {1},
  pages = {93},
  year = {2025},
  month = {Aug},
  publisher = {Nature Publishing Group},
  doi = {10.1038/s41535-025-00804-0},
  url = {https://www.nature.com/articles/s41535-025-00804-0}
}

@article{Eremin12,
  title = {Dual features of magnetic susceptibility in superconducting cuprates: a comparison to inelastic neutron scattering},
  author = {Eremin, M. V. and Shigapov, I. M. and Eremin, I. M.},
  journal = {The European Physical Journal B},
  volume = {85},
  pages = {131},
  year = {2012},
  publisher = {Springer},
  doi = {10.1140/epjb/e2012-20539-y},
  url = {https://link.springer.com/article/10.1140/epjb/e2012-20539-y}
}

@article{Reznik08,
  title = {Local moment fluctuations in an optimally-doped high-{$T_c$} superconductor},
  author = {Reznik, D. and Ismer, J.-P. and Eremin, I. and Pintschovius, L. and Arai, M. and Endoh, Y. and Masui, T. and Tajima, S.},
  journal = {Physical Review B},
  volume = {78},
  number = {13},
  pages = {132503},
  year = {2008},
  month = {Oct},
  publisher = {American Physical Society},
  doi = {10.1103/PhysRevB.78.132503},
  url = {https://link.aps.org/doi/10.1103/PhysRevB.78.132503}
}

@article{Hinkov07,
  title = {Spin dynamics in the pseudogap state of a high-temperature superconductor},
  author = {Hinkov, V. and Bourges, P. and Pailh{\`e}s, S. and Sidis, Y. and Ivanov, A. and Frost, C. D. and Perring, T. G. and Lin, C. T. and Chen, D. P. and Keimer, B.},
  journal = {Nature Physics},
  volume = {3},
  number = {11},
  pages = {780},
  year = {2007},
  month = {Nov},
  publisher = {Nature Publishing Group},
  doi = {10.1038/nphys720},
  url = {https://www.nature.com/articles/nphys720}
}

@article{Chan16,
  title = {Hourglass Dispersion and Resonance of Magnetic Excitations in the Superconducting State of the Single-Layer Cuprate {${\mathrm{HgBa}}_{2}{\mathrm{CuO}}_{4+\ensuremath{\delta}}$} Near Optimal Doping},
  author = {Chan, M. K. and Tang, Y. and Dorow, C. J. and Jeong, J. and Mangin-Thro, L. and Veit, M. J. and Ge, Y. and Abernathy, D. L. and Sidis, Y. and Bourges, P. and Greven, M.},
  journal = {Phys. Rev. Lett.},
  volume = {117},
  issue = {27},
  pages = {277002},
  numpages = {6},
  year = {2016},
  month = {Dec},
  publisher = {American Physical Society},
  doi = {10.1103/PhysRevLett.117.277002},
  url = {https://link.aps.org/doi/10.1103/PhysRevLett.117.277002}
}

@article{Chan16_1,
  title = {Commensurate antiferromagnetic excitations as a signature of the pseudogap in the tetragonal high-{$T_c$} cuprate {$HgBa_2CuO_{4+\delta}$}},
  author = {Chan, M. K. and Dorow, C. J. and Mangin-Thro, L. and Tang, Y. and Ge, Y. and Veit, M. J. and Yu, G. and Zhao, X. and Christianson, A. D. and Park, J. T. and Sidis, Y. and Steffens, P. and Abernathy, D. L. and Bourges, P. and Greven, M.},
  journal = {Nature Communications},
  volume = {7},
  number = {1},
  pages = {10819},
  year = {2016},
  month = {Mar},
  publisher = {Nature Publishing Group},
  doi = {10.1038/ncomms10819},
  url = {https://www.nature.com}
}

@article{Rohringer18,
  title = {Diagrammatic routes to nonlocal correlations between electrons in solids},
  author = {Rohringer, G. and Hafermann, H. and Toschi, A. and Katanin, A. A. and Antipov, A. E. and Katsnelson, M. I. and Lichtenstein, A. I. and Rubtsov, A. N. and Held, K.},
  journal = {Reviews of Modern Physics},
  volume = {90},
  number = {2},
  pages = {025003},
  year = {2018},
  month = {Jun},
  publisher = {American Physical Society},
  doi = {10.1103/RevModPhys.90.025003},
  url = {https://link.aps.org}
}

@article{Maier05,
  title = {Quantum cluster approximations},
  author = {Maier, Thomas and Jarrell, Mark and Pruschke, Thomas and Hettler, Matthias H.},
  journal = {Reviews of Modern Physics},
  volume = {77},
  number = {3},
  pages = {1027},
  year = {2005},
  month = {Oct},
  publisher = {American Physical Society},
  doi = {10.1103/RevModPhys.77.1027},
  url = {https://link.aps.org}
}

@article{Pelc20,
  title = {Resistivity phase diagram of cuprates revisited},
  author = {Pelc, D. and Veit, M. J. and Dorow, C. J. and Ge, Y. and Bari{\v{s}}i{\'c}, N. and Greven, M.},
  journal = {Physical Review B},
  volume = {102},
  number = {7},
  pages = {075114},
  year = {2020},
  month = {Aug},
  publisher = {American Physical Society},
  doi = {10.1103/PhysRevB.102.075114},
  url = {https://link.aps.org/doi/10.1103/PhysRevB.102.075114}
}

@book{Ovchinnikov_book,
  title = {{H}ubbard Operators in the Theory of Strongly Correlated Electrons},
  author = {Ovchinnikov, S. G. and Val'kov, V. V.},
  year = {2004},
  publisher = {Imperial College Press},
  address = {London},
  doi = {10.1142/9781860946059},
  url = {https://www.worldscientific.com/worldscibooks/10.1142/p314?srsltid=AfmBOorE2z27kQsA0f-JQHXR17coqr5Tn8lIpatGd4BM6muaz6KuIqro#t=aboutBook},
  isbn = {978-1-86094-442-0}
}

@article{Wang20,
  title={Emergence of quasiparticles in a doped {Mott} insulator},
  author={Wang, Yao and He, Yu and Wohlfeld, Krzysztof and Hashimoto, Makoto and Huang, Edwin W and Lu, Donghui and Mo, Sung-Kwan and Komiya, Seiki and Jia, Chunjing and Moritz, Brian and others},
  journal={Communications Physics},
  volume={3},
  number={1},
  pages={210},
  year={2020},
  doi = {10.1038/s42005-020-00480-5},
  url = {https://doi.org/10.1038/s42005-020-00480-5},
  publisher={Nature Publishing Group UK London}
}

@article{Senechal00,
  title = {Spectral {W}eight of the {H}ubbard Model through {C}luster {P}erturbation {T}heory},
  author = {S\'en\'echal, D. and Perez, D. and Pioro-Ladri\`ere, M.},
  journal = {Phys. Rev. Lett.},
  volume = {84},
  issue = {3},
  pages = {522},
  numpages = {0},
  year = {2000},
  month = {Jan},
  publisher = {American Physical Society},
  doi = {10.1103/PhysRevLett.84.522},
  url = {https://link.aps.org/doi/10.1103/PhysRevLett.84.522}
}

@article{Senechal02,
  title = {Cluster perturbation theory for {H}ubbard models},
  author = {S\'en\'echal, David and Perez, Danny and Plouffe, Dany},
  journal = {Phys. Rev. B},
  volume = {66},
  issue = {7},
  pages = {075129},
  numpages = {11},
  year = {2002},
  month = {Aug},
  publisher = {American Physical Society},
  doi = {10.1103/PhysRevB.66.075129},
  url = {https://link.aps.org/doi/10.1103/PhysRevB.66.075129}
}

@article{Senechal04,
  title = {Hot {S}pots and {P}seudogaps for {H}ole- and {E}lectron-{D}oped {H}igh-{T}emperature {S}uperconductors},
  author = {S\'en\'echal, David and Tremblay, A.-M. S.},
  journal = {Phys. Rev. Lett.},
  volume = {92},
  issue = {12},
  pages = {126401},
  numpages = {4},
  year = {2004},
  month = {Mar},
  publisher = {American Physical Society},
  doi = {10.1103/PhysRevLett.92.126401},
  url = {https://link.aps.org/doi/10.1103/PhysRevLett.92.126401}
}

@article{Seki18,
  title = {Variational cluster approach to thermodynamic properties of interacting fermions at finite temperatures: A case study of the two-dimensional single-band {H}ubbard model at half filling},
  author = {Seki, Kazuhiro and Shirakawa, Tomonori and Yunoki, Seiji},
  journal = {Phys. Rev. B},
  volume = {98},
  issue = {20},
  pages = {205114},
  numpages = {37},
  year = {2018},
  month = {Nov},
  publisher = {American Physical Society},
  doi = {10.1103/PhysRevB.98.205114},
  url = {https://link.aps.org/doi/10.1103/PhysRevB.98.205114}
}

@article{Kuzmin20,
  title = {Doping and temperature evolution of pseudogap and spin-spin correlations in the two-dimensional {H}ubbard model},
  author = {Kuz'min, V. I. and Visotin, M. A. and Nikolaev, S. V. and Ovchinnikov, S. G.},
  journal = {Phys. Rev. B},
  volume = {101},
  issue = {11},
  pages = {115141},
  numpages = {12},
  year = {2020},
  month = {Mar},
  publisher = {American Physical Society},
  doi = {10.1103/PhysRevB.101.115141},
  url = {https://link.aps.org/doi/10.1103/PhysRevB.101.115141}
}

@article{Plakida13,
  title={On the theory of superconductivity in the extended {H}ubbard model: Spin-fluctuation pairing},
  author={Plakida, Nikolay M and Oudovenko, Viktor S},
  journal={The European Physical Journal B},
  volume={86},
  pages={1},
  year={2013},
  publisher={Springer},
  doi = {10.1140/epjb/e2013-31157-6},
  url={https://doi.org/10.1140/epjb/e2013-31157-6}
}

@article{Zubarev60,
  title={Double-time {G}reen functions in statistical physics},
  author={Zubarev, D. N.},
  journal={Soviet Physics Uspekhi},
  volume={3},
  number={3},
  pages={320},
  year={1960},
  publisher={IOP Publishing},
  doi={10.1070/PU1960v003n03ABEH003275},
  url={https://doi.org/10.1070/PU1960v003n03ABEH003275}
}

@article{Mori65,
  title={A continued-fraction representation of the time-correlation functions},
  author={Mori, Hazime},
  journal={Progress of Theoretical Physics},
  volume={34},
  number={3},
  pages={399},
  year={1965},
  publisher={Oxford University Press},
  doi={10.1143/PTP.34.399},
  url={https://doi.org/10.1143/PTP.34.399}
}

@article{Korshunov07,
  title={Doping-dependent evolution of low-energy excitations and quantum phase transitions within an effective model for high-{T}c copper oxides},
  author={Korshunov, MM and Ovchinnikov, SG},
  journal={The European Physical Journal B},
  volume={57},
  pages={271},
  year={2007},
  publisher={Springer},
  doi={10.1140/epjb/e2007-00179-2},
  url={https://doi.org/10.1140/epjb/e2007-00179-2}
}

@article{Shneyder20,
  title = {Polaron transformations in the realistic model of the strongly correlated electron system},
  author = {Shneyder, E. I. and Nikolaev, S. V. and Zotova, M. V. and Kaldin, R. A. and Ovchinnikov, S. G.},
  journal = {Phys. Rev. B},
  volume = {101},
  issue = {23},
  pages = {235114},
  numpages = {14},
  year = {2020},
  month = {Jun},
  publisher = {American Physical Society},
  doi = {10.1103/PhysRevB.101.235114},
  url = {https://link.aps.org/doi/10.1103/PhysRevB.101.235114}
}

@article{Werner09,
  title = {Momentum-sector-selective metal-insulator transition in the eight-site dynamical mean-field approximation to the {H}ubbard model},
  author = {Werner, Philipp and Gull, Emanuel and Parcollet, Olivier and Millis, Andrew J.},
  journal = {Physical Review B},
  volume = {80},
  number = {4},
  pages = {045120},
  year = {2009},
  month = {Jul},
  publisher = {American Physical Society},
  doi = {10.1103/PhysRevB.80.045120},
  url = {https://link.aps.org}
}

@article{Gull10,
  title = {Momentum-space anisotropy and pseudogaps: A comparative cluster dynamical mean-field analysis of the doping-driven metal-insulator transition in the two-dimensional {H}ubbard model},
  author = {Gull, E. and Ferrero, M. and Parcollet, O. and Georges, A. and Millis, A. J.},
  journal = {Phys. Rev. B},
  volume = {82},
  issue = {15},
  pages = {155101},
  numpages = {14},
  year = {2010},
  month = {Oct},
  publisher = {American Physical Society},
  doi = {10.1103/PhysRevB.82.155101},
  url = {https://link.aps.org/doi/10.1103/PhysRevB.82.155101}
}

@article{Timusk99,
  title={The pseudogap in high-temperature superconductors: an experimental survey},
  author={Timusk, Tom and Statt, Bryan},
  journal={Reports on Progress in Physics},
  volume={62},
  number={1},
  pages={61},
  year={1999},
  publisher={IOP Publishing},
  doi={10.1088/0034-4885/62/1/002},
  url={https://iopscience.iop.org/article/10.1088/0034-4885/62/1/002}
}

@article{Robinson19,
  title={Anomalies in the pseudogap phase of the cuprates: competing ground states and the role of umklapp scattering},
  author={Robinson, Neil J and Johnson, Peter D and Rice, T Maurice and Tsvelik, Alexei M},
  journal={Reports on Progress in Physics},
  volume={82},
  number={12},
  pages={126501},
  year={2019},
  publisher={IOP Publishing},
  doi={10.1088/1361-6633/ab31ed},
  url={https://iopscience.iop.org/article/10.1088/1361-6633/ab31ed}
}

@article{Sadovskii05,
  title = {Pseudogaps in strongly correlated metals: A generalized dynamical mean-field theory approach},
  author = {Sadovskii, M. V. and Nekrasov, I. A. and Kuchinskii, E. Z. and Pruschke, Th. and Anisimov, V. I.},
  journal = {Phys. Rev. B},
  volume = {72},
  issue = {15},
  pages = {155105},
  numpages = {11},
  year = {2005},
  month = {Oct},
  publisher = {American Physical Society},
  doi = {10.1103/PhysRevB.72.155105},
  url = {https://link.aps.org/doi/10.1103/PhysRevB.72.155105}
}

@article{Metzner89,
  title = {Correlated Lattice Fermions in $d=\ensuremath{\infty}$ Dimensions},
  author = {Metzner, Walter and Vollhardt, Dieter},
  journal = {Phys. Rev. Lett.},
  volume = {62},
  issue = {3},
  pages = {324},
  numpages = {0},
  year = {1989},
  month = {Jan},
  publisher = {American Physical Society},
  doi = {10.1103/PhysRevLett.62.324},
  url = {https://link.aps.org/doi/10.1103/PhysRevLett.62.324}
}

@article{Georges96,
  title = {Dynamical mean-field theory of strongly correlated fermion systems and the limit of infinite dimensions},
  author = {Georges, Antoine and Kotliar, Gabriel and Krauth, Werner and Rozenberg, Marcelo J.},
  journal = {Rev. Mod. Phys.},
  volume = {68},
  issue = {1},
  pages = {13},
  numpages = {0},
  year = {1996},
  month = {Jan},
  publisher = {American Physical Society},
  doi = {10.1103/RevModPhys.68.13},
  url = {https://link.aps.org/doi/10.1103/RevModPhys.68.13}
}

@article{Zaitsev75,
  author  = {Zaitsev, R. O.},
  title   = {Generalized diagram technique and spin waves in an anisotropic ferromagnet},
  journal = {Soviet Physics JETP},
  year    = {1975},
  volume  = {41},
  number  = {1},
  pages   = {100},
  url     = {https://www.jetp.ras.ru/cgi-bin/e/index/e/41/1/p100?a=list},
  note    = {English translation of Zh. Eksp. Teor. Fiz. 68, 207 (1975)}
}

@article{Kuzmin24,
  author  = {Kuz'min, V. I. and Korshunov, M. M. and Nikolaev, S. V. and Ovchinnikova, T. M. and Ovchinnikov, S. G.},
  title   = {Interrelation between Doping Dependencies of the Spin Susceptibility and Electronic Structure in Cuprates},
  journal = {JETP Letters},
  year    = {2024},
  volume  = {120},
  number  = {1},
  pages   = {46},
  doi     = {10.1134/S0021364024601945},
  url     = {https://link.springer.com/article/10.1134/S0021364024601945}
}

@article{Kruger07,
  title = {Magnetic fluctuations in $n$-type high-${T}_{c}$ superconductors reveal breakdown of fermiology: {Experiments} and {Fermi}-liquid/{RPA} calculations},
  author = {Kr\"uger, F. and Wilson, S. D. and Shan, L. and Li, Shiliang and Huang, Y. and Wen, H.-H. and Zhang, S.-C. and Dai, Pengcheng and Zaanen, J.},
  journal = {Phys. Rev. B},
  volume = {76},
  issue = {9},
  pages = {094506},
  numpages = {9},
  year = {2007},
  month = {Sep},
  publisher = {American Physical Society},
  doi = {10.1103/PhysRevB.76.094506},
  url = {https://link.aps.org/doi/10.1103/PhysRevB.76.094506}
}

@article{Wang25,
  title = {Probing the pseudogap and beyond: examining single-particle properties of the hole- and electron-doped {Hubbard} model},
  author = {Wang, Wen O. and Huang, Edwin W. and Moritz, Brian and Devereaux, Thomas P.},
  journal = {arXiv preprint arXiv:2506.15770},
  year = {2025},
  url = {https://arxiv.org/abs/2506.15770}
}

@article{Yamaji11,
  title = {Composite-Fermion Theory for Pseudogap, {Fermi} Arc, Hole Pocket, and Non-{Fermi} Liquid of Underdoped Cuprate Superconductors},
  author = {Yamaji, Youhei and Imada, Masatoshi},
  journal = {Phys. Rev. Lett.},
  volume = {106},
  issue = {1},
  pages = {016404},
  numpages = {4},
  year = {2011},
  month = {Jan},
  publisher = {American Physical Society},
  doi = {10.1103/PhysRevLett.106.016404},
  url = {https://link.aps.org/doi/10.1103/PhysRevLett.106.016404}
}

@article{Sakai16,
  title = {Hidden-fermion representation of self-energy in pseudogap and superconducting states of the two-dimensional {Hubbard} model},
  author = {Sakai, Shiro and Civelli, Marcello and Imada, Masatoshi},
  journal = {Phys. Rev. B},
  volume = {94},
  issue = {11},
  pages = {115130},
  numpages = {14},
  year = {2016},
  month = {Sep},
  publisher = {American Physical Society},
  doi = {10.1103/PhysRevB.94.115130},
  url = {https://link.aps.org/doi/10.1103/PhysRevB.94.115130}
}

@article{Plakida07,
  title={Electron spectrum in high-temperature cuprate superconductors},
  author={Plakida, NM and Oudovenko, VS},
  journal={Journal of Experimental and Theoretical Physics},
  volume={104},
  number={2},
  pages={230},
  year={2007},
  publisher={Springer},
  doi={10.1134/S1063776107020082},
  url={https://link.springer.com/article/10.1134/S1063776107020082}
}

@article{Tung21,
title = {Electronic spectrum and superconductivity in the extended {t–J–V} model},
journal = {Physica C: Superconductivity and its Applications},
volume = {587},
pages = {1353900},
year = {2021},
issn = {0921-4534},
doi = {https://doi.org/10.1016/j.physc.2021.1353900},
url = {https://www.sciencedirect.com/science/article/pii/S0921453421000836},
author = {Nguen {Dan Tung} and Artem A. Vladimirov and Nikolay M. Plakida},
}

@article{Belemuk16,
  title = {On the spectral function of carriers in the pseudogap state},
  author = {Belemuk, A. M. and Barabanov, A. F.},
  journal = {Journal of Experimental and Theoretical Physics},
  volume = {123},
  number = {3},
  pages = {470},
  year = {2016},
  publisher = {Springer},
  doi = {10.1134/S1063776116070141},
  url = {https://link.springer.com/article/10.1134/S1063776116070141}
}

@article{Gull08,
  title = {Superconducting and pseudogap phases of the {H}ubbard model: A cluster-dynamical mean-field study},
  author = {Gull, E. and Werner, P. and Wang, X. and Troyer, M. and Millis, A. J.},
  journal = {Europhysics Letters},
  volume = {84},
  number = {3},
  pages = {37009},
  year = {2008},
  doi = {10.1209/0295-5075/84/37009},
  url = {https://iopscience.iop.org}
}

@article{Gull09,
  title = {Momentum-sector-selective metal-insulator transition in the eight-site dynamical mean-field approximation to the {Hubbard} model in two dimensions},
  author = {Gull, Emanuel and Parcollet, Olivier and Werner, Philipp and Millis, Andrew J.},
  journal = {Phys. Rev. B},
  volume = {80},
  issue = {24},
  pages = {245102},
  numpages = {13},
  year = {2009},
  month = {Dec},
  publisher = {American Physical Society},
  doi = {10.1103/PhysRevB.80.245102},
  url = {https://link.aps.org/doi/10.1103/PhysRevB.80.245102}
}

@Article{Kuzmin23,
AUTHOR = {Kuz’min, Valerii I. and Nikolaev, Sergey V. and Korshunov, Maxim M. and Ovchinnikov, Sergey G.},
TITLE = {One- and Two-Particle Correlation Functions in the Cluster Perturbation Theory for Cuprates},
JOURNAL = {Materials},
VOLUME = {16},
YEAR = {2023},
NUMBER = {13},
pages = {4640},
URL = {https://www.mdpi.com/1996-1944/16/13/4640},
PubMedID = {37444953},
ISSN = {1996-1944},
DOI = {10.3390/ma16134640}
}

@article{Chao77,
doi = {10.1088/0022-3719/10/10/002},
url = {https://doi.org/10.1088/0022-3719/10/10/002},
year = {1977},
month = {may},
publisher = {},
volume = {10},
number = {10},
pages = {L271},
author = {K A Chao and J Spalek and A M Oles},
title = {Kinetic exchange interaction in a narrow S-band},
journal = {Journal of Physics C: Solid State Physics}
}

@article{Mendels88,
title = {Zero field {NMR} of the magnetic copper sites in antiferromagnetic $\text{YBa}_2\text{Cu}_3\text{O}_{6+x}$},
journal = {Physica C: Superconductivity},
volume = {156},
number = {3},
pages = {355-358},
year = {1988},
issn = {0921-4534},
doi = {https://doi.org/10.1016/0921-4534(88)90757-5},
url = {https://www.sciencedirect.com/science/article/pii/0921453488907575},
author = {P. Mendels and H. Alloul}
}

@article{Friedel89,
doi = {10.1088/0953-8984/1/42/001},
url = {https://doi.org/10.1088/0953-8984/1/42/001},
year = {1989},
month = {oct},
publisher = {},
volume = {1},
number = {42},
pages = {7757},
author = {J Friedel},
title = {The high-{Tc} superconductors: a conservative view},
journal = {Journal of Physics: Condensed Matter}
}

@article{Zhu23,
  title = {Spin fluctuations associated with the collapse of the pseudogap in a cuprate superconductor},
  author = {Zhu, M. and Voneshen, D. J. and Raymond, S. and Lipscombe, O. J. and Tam, C. C. and Hayden, S. M.},
  journal = {Nat. Phys.},
  volume = {19},
  issue = {1},
  pages = {99--105},
  numpages = {7},
  year = {2023},
  month = {Jan},
  publisher = {Nature Publishing Group},
  doi = {10.1038/s41567-022-01825-3},
  url = {https://doi.org}
}

@article{Alloul89,
  title = {$^{89}\mathrm{Y}$ {NMR} evidence for a {Fermi}-liquid behavior in $\text{YBa}_2\text{Cu}_3\text{O}_{6+x}$},
  author = {Alloul, H. and Ohno, T. and Mendels, P.},
  journal = {Phys. Rev. Lett.},
  volume = {63},
  issue = {16},
  pages = {1700--1703},
  numpages = {0},
  year = {1989},
  month = {Oct},
  publisher = {American Physical Society},
  doi = {10.1103/PhysRevLett.63.1700},
  url = {https://link.aps.org/doi/10.1103/PhysRevLett.63.1700}
}

@article{Alloul12,
  author    = {Alloul, Henri},
  title     = {From Friedel Oscillations and {Kondo} Effect to the Pseudogap in Cuprates},
  journal   = {Journal of Superconductivity and Novel Magnetism},
  year      = {2012},
  volume    = {25},
  number    = {3},
  pages     = {585},
  doi       = {10.1007/s10948-012-1472-x},
  url       = {https://doi.org/10.1007/s10948-012-1472-x}
}

@article{Barzykin95,
  title = {Magnetic scaling in cuprate superconductors},
  author = {Barzykin, V. and Pines, D.},
  journal = {Phys. Rev. B},
  volume = {52},
  issue = {18},
  pages = {13585--13600},
  numpages = {0},
  year = {1995},
  month = {Nov},
  publisher = {American Physical Society},
  doi = {10.1103/PhysRevB.52.13585},
  url = {https://link.aps.org/doi/10.1103/PhysRevB.52.13585}
}

@article{Kordyuk15,
    author = {Kordyuk, A. A.},
    title = {Pseudogap from ARPES experiment: Three gaps in cuprates and topological superconductivity (Review Article)},
    journal = {Low Temperature Physics},
    volume = {41},
    number = {5},
    pages = {319-341},
    year = {2015},
    month = {05},
    issn = {1063-777X},
    doi = {10.1063/1.4919371},
    url = {https://doi.org/10.1063/1.4919371}
}

@article{Loeser96,
author = {A. G. Loeser  and Z.-X. Shen  and D. S. Dessau  and D. S. Marshall  and C. H. Park  and P. Fournier  and A. Kapitulnik },
title = {Excitation Gap in the Normal State of Underdoped ${Bi_2Sr_2CaCu_2O}_{8+\delta}$},
journal = {Science},
volume = {273},
number = {5273},
pages = {325-329},
year = {1996},
doi = {10.1126/science.273.5273.325},
URL = {https://www.science.org/doi/abs/10.1126/science.273.5273.325},
}

@article{Schmalian98,
  title = {Weak Pseudogap Behavior in the Underdoped Cuprate Superconductors},
  author = {Schmalian, J\"org and Pines, David and Stojkovi\ifmmode \acute{c}\else \'{c}\fi{}, Branko},
  journal = {Phys. Rev. Lett.},
  volume = {80},
  issue = {17},
  pages = {3839--3842},
  numpages = {0},
  year = {1998},
  month = {Apr},
  publisher = {American Physical Society},
  doi = {10.1103/PhysRevLett.80.3839},
  url = {https://link.aps.org/doi/10.1103/PhysRevLett.80.3839}
}

@article{Chubukov98,
  title = {Temperature variation of the pseudogap in underdoped cuprates},
  author = {Chubukov, Andrey V. and Schmalian, J\"org},
  journal = {Phys. Rev. B},
  volume = {57},
  issue = {18},
  pages = {R11085(R)--R11088(R)},
  numpages = {0},
  year = {1998},
  month = {May},
  publisher = {American Physical Society},
  doi = {10.1103/PhysRevB.57.R11085},
  url = {https://link.aps.org/doi/10.1103/PhysRevB.57.R11085}
}

@article{Chubukov98_1,
	doi = {10.1209/epl/i1998-00522-9},
	url = {https://doi.org},
	year = {1998},
	month = {dec},
	publisher = {{IOP} Publishing},
	volume = {44},
	number = {5},
	pages = {655--661},
	author = {A. V. Chubukov},
	title = {Theory of the leading edge gap in underdoped cuprates},
	journal = {Europhysics Letters ({EPL})}
}

@article{Millis90,
  title = {Phenomenological model of nuclear relaxation in the normal state of $\text{YBa}_2\text{Cu}_3\text{O}_{7}$},
  author = {Millis, A. J. and Monien, Hartmut and Pines, David},
  journal = {Phys. Rev. B},
  volume = {42},
  issue = {1},
  pages = {167--178},
  numpages = {0},
  year = {1990},
  month = {Jul},
  publisher = {American Physical Society},
  doi = {10.1103/PhysRevB.42.167},
  url = {https://link.aps.org/doi/10.1103/PhysRevB.42.167}
}

@article{Schmalian99,
  title = {Microscopic theory of weak pseudogap behavior in the underdoped cuprate superconductors: General theory and quasiparticle properties},
  author = {Schmalian, J\"org and Pines, David and Stojkovi\ifmmode \acute{c}\else \'{c}\fi{}, Branko},
  journal = {Phys. Rev. B},
  volume = {60},
  issue = {1},
  pages = {667--686},
  numpages = {0},
  year = {1999},
  month = {Jul},
  publisher = {American Physical Society},
  doi = {10.1103/PhysRevB.60.667},
  url = {https://link.aps.org/doi/10.1103/PhysRevB.60.667}
}

@article{Vladimirov09,
  title = {Dynamic spin susceptibility in the t-{J} model},
  author = {Vladimirov, A. A. and Ihle, D. and Plakida, N. M.},
  journal = {Phys. Rev. B},
  volume = {80},
  issue = {10},
  pages = {104425},
  numpages = {12},
  year = {2009},
  month = {Sep},
  publisher = {American Physical Society},
  doi = {10.1103/PhysRevB.80.104425},
  url = {https://link.aps.org/doi/10.1103/PhysRevB.80.104425}
}

\end{document}